\documentclass[twoside,twocolumn,9pt]{article}
\usepackage{extsizes}
\usepackage[super,sort&compress,comma]{natbib} 
\usepackage[version=3]{mhchem}
\usepackage[left=1.5cm, right=1.5cm, top=1.785cm, bottom=2.0cm]{geometry}
\usepackage{balance}
\usepackage{mathptmx}
\usepackage{sectsty}
\usepackage{graphicx} 
\usepackage{subcaption}
\usepackage{lastpage}
\usepackage[format=plain,justification=justified,singlelinecheck=false,font={stretch=1.125,small,sf},labelfont=bf,labelsep=space]{caption}
\usepackage{float}
\usepackage{fancyhdr}
\usepackage{fnpos}
\usepackage[english]{babel}
\addto{\captionsenglish}{%
  
}
\usepackage{array}
\usepackage{droidsans}
\usepackage{charter}
\usepackage[T1]{fontenc}
\usepackage[usenames,dvipsnames]{xcolor}
\usepackage{setspace}
\usepackage[compact]{titlesec}
\usepackage{hyperref}
\usepackage{amsmath}
\usepackage{amssymb}
\usepackage{comment}

\usepackage{mathtools}
\usepackage{multirow}
\usepackage{multicol}
\newcommand{\xdownarrow}[1]{%
  {\left\downarrow\vbox to #1{}\right.\kern-\nulldelimiterspace}
}
\usepackage{siunitx}
\usepackage{enumitem}

\definecolor{cream}{RGB}{222,217,201}

\begin{document}

\pagestyle{fancy}
\thispagestyle{plain}
\fancypagestyle{plain}{
\renewcommand{\headrulewidth}{0pt}
}

\makeFNbottom
\makeatletter
\renewcommand\LARGE{\@setfontsize\LARGE{15pt}{17}}
\renewcommand\Large{\@setfontsize\Large{12pt}{14}}
\renewcommand\large{\@setfontsize\large{10pt}{12}}
\renewcommand\footnotesize{\@setfontsize\footnotesize{7pt}{10}}
\makeatother

\renewcommand{\thefootnote}{\fnsymbol{footnote}}
\renewcommand\footnoterule{\vspace*{1pt}%
\color{cream}\hrule width 3.5in height 0.4pt \color{black}\vspace*{5pt}} 
\setcounter{secnumdepth}{5}

\makeatletter 
\renewcommand\@biblabel[1]{#1}            
\renewcommand\@makefntext[1]%
{\noindent\makebox[0pt][r]{\@thefnmark\,}#1}
\makeatother 
\renewcommand{\figurename}{\small{Fig.}~}
\sectionfont{\sffamily\Large}
\subsectionfont{\normalsize}
\subsubsectionfont{\bf}
\setstretch{1.125} 
\setlength{\skip\footins}{0.8cm}
\setlength{\footnotesep}{0.25cm}
\setlength{\jot}{10pt}
\titlespacing*{\section}{0pt}{4pt}{4pt}
\titlespacing*{\subsection}{0pt}{15pt}{1pt}

\fancyfoot{}
\fancyfoot[RO]{\footnotesize{\sffamily{1--\pageref{LastPage} ~\textbar  \hspace{2pt}\thepage}}}
\fancyfoot[LE]{\footnotesize{\sffamily{\thepage~\textbar\hspace{4.65cm} 1--\pageref{LastPage}}}}
\fancyhead{}
\renewcommand{\headrulewidth}{0pt} 
\renewcommand{\footrulewidth}{0pt}
\setlength{\arrayrulewidth}{1pt}
\setlength{\columnsep}{6.5mm}
\setlength\bibsep{1pt}

\makeatletter 
\newlength{\figrulesep} 
\setlength{\figrulesep}{0.5\textfloatsep} 

\newcommand{\topfigrule}{\vspace*{-1pt}%
\noindent{\color{black}\rule[-\figrulesep]{\columnwidth}{1.5pt}} }

\newcommand{\botfigrule}{\vspace*{-2pt}%
\noindent{\color{black}\rule[\figrulesep]{\columnwidth}{1.5pt}} }

\newcommand{\dblfigrule}{\vspace*{-1pt}%
\noindent{\color{black}\rule[-\figrulesep]{\textwidth}{1.5pt}} }

\makeatother

\twocolumn[
  \begin{@twocolumnfalse}
{
}\par
\vspace{1em}
\sffamily

{Preprint}\\
\vskip -3mm
 \noindent\LARGE{\textbf{Free energy differences and coexistence of clathrate structures II and H via~lattice-switch~Monte Carlo}} \\
\vskip -3mm
  \noindent\large{Olivia S. Moro,$^{\ast}$\textit{$^{ab}$}, Nigel B. Wilding\textit{$^{b}$$^{\ddag}$} and Vincent Ballenegger\textit{$^{a}$}.} \\
\vskip 5mm
\centerline{
\parbox{0.75\linewidth}{
\noindent\normalsize{We introduce a simulation technique to compute the free energy difference between two hydrate structures of different stoichiometry connected to a reservoir of gas molecules at a prescribed pressure. The method permits the determination of coexistence parameters for the system when the two hydrate structures have the same number of water molecules $N_w$. The approach is based on performing isobaric Lattice Switch Monte Carlo simulations to measure free energy differences between the hydrate structures when they are either fully occupied by gas molecules, or fully empty. This measurement is combined with thermodynamic integration within an ensemble in which the number of guest molecules $N_g$ can fluctuate under the control of a chemical potential $\mu_g$. We analyze the properties of the resulting constant-$N_w,\mu_g,P,T$ ensemble and show how it can be used to calculate coexistence points via a thermodynamic cycle. Applying the method to argon and methane structures, we find coexistence pressures that are in good agreement overall with the available experimental data. }}
}

\bigskip



 \end{@twocolumnfalse} \vspace{1cm}
  ]

\renewcommand*\rmdefault{bch}\normalfont\upshape
\rmfamily
\section*{}
\vspace{-1cm}


\footnotetext{\textit{$^{a}$~ Université Marie et Louis Pasteur, CNRS, Institut UTINAM (UMR 6213), F-25000 Besançon, France}}
\footnotetext{\textit{$^{b}$~HH Wills Physics Laboratory, Royal Fort, University of Bristol, Bristol BS8 1TL, U.K. }}
\footnotetext{\textit{$^{\ast}$~ Email: hn22404@bristol.ac.uk}}
\footnotetext{\textit{$^{\ddag}$~ Email: nigel.wilding@bristol.ac.uk}}




\section{Introduction}

A mixture of water and gas (e.g. CH$_{4}$, Ar, CO$_2$) at sufficiently low temperatures and/or high pressures forms spontaneously a gas hydrate, an ice-like crystalline structure in which guest gas molecules are trapped (``enclathrated'') inside of hydrogen-bonded water cages. This inclusion compound, also known as a clathrate hydrate, is stabilized by van der Waals interactions between the guest molecules and the water framework. It is relevant in a variety of scientific and industrial contexts, including flow assurance in pipelines, gas separation, gas storage, carbon dioxide sequestration, and planetary science. \cite{HydrateReview} 

The most common clathrate structures are structure I (sI), II (sII), and H (sH).~\cite{HydrateReview} Structures I and II contain two cavity types: small and large cages. At low to moderate pressures, small gaseous molecules (Ar, N$_2$, ...) favor sII because it contains about three times more small cages than sI. Structure H has three cavity types, the largest one being $\sim$20\% larger than the large cavities in sI and sII. It can form in the presence of large guest molecules that fit only in the large cages of sH, but also more generally at high pressures because the large cages can accommodate easily several small molecules~\cite{Loveday2008}. A transition to sH has been observed in pure gas hydrates of Ar, Xe, Kr, N$_2$ and CH$_4$ at a high pressure typically between 0.5 and 1 GPa.\cite{Loveday2008,Hirai2023} Other phases can become more stable at even higher pressures, like the tetragonal structure (sT) or a filled-ice structure.~\cite{Loveday2008, Hirai2023,Tulk2014}

The statistical mechanical model of van der Waals and Platteeuw has often been used to predict the stability domain (dissociation point) and composition of a gas hydrate as a function of temperature, pressure, and composition of the gas phase (pure or mixed)~\cite{vdW1959,Punnathanam}. The approximate and semiempirical nature of this model limits however, its usefulness, especially for atypical conditions such as high pressures encountered in planetary science.\cite{Hirai2023} 

The direct computation of the phase diagram of hydrates, using molecular simulations with intermolecular potentials, was pioneered by Wierzchowski and Monson (2006) \cite{Wierzchowski} and Jensen et al. \cite{Jensen2010}. In these studies, rather intricate free energy calculations are carried out to determine the chemical potentials of water and gas in the three relevant phases—liquid water, gas hydrate, and free gas—as functions of temperature at a constant pressure. The dissociation point, where these three phases coexist, corresponds to the temperature at which their chemical potentials become equal. The calculations involve an intermediate reference state, the empty clathrate, for which the chemical potential of water must be determined before subsequently filling the structure with the gas of interest. To obtain this chemical potential, the Frenkel-Ladd method \cite{FrenkelLadd1984} is employed, using another reference state with a known analytical free energy: an Einstein molecular crystal, in which water molecules are tethered to their ideal positions and orientations via harmonic potentials. This approach has been applied notably to map the stability domain of sI gas hydrates containing CH$_4$ and CO$_2$ \cite{Waage}.

An alternative approach considers direct simulations of phase coexistence. Although conceptually straightforward, this method is computationally more intensive than the previously described free energy method. To determine the dissociation point of a gas hydrate, long molecular dynamics simulations are conducted within the isothermal-isobaric ensemble (constant \(N_w, N_g, P, T\)) on a system that includes gas hydrate, liquid water, and free gas molecules. The coexistence point is identified as the temperature midpoint between the highest temperature at which the hydrate forms and the lowest temperature at which it dissociates. Jin and Coasne~\cite{CoasneJin2017} applied this technique using the grand-canonical ensemble (constant \(\mu_w, \mu_g, V, T\)) and obtained a methane hydrate dissociation temperature that closely aligned with results from free energy calculations~\cite{Waage,CoasneJin2017}. Implementing the direct coexistence method within the grand-canonical ensemble is advantageous, as it allows the number of gas molecules to vary dynamically during hydrate growth. 

\begin{figure}
    \centering
    \includegraphics[width=1.0\linewidth]{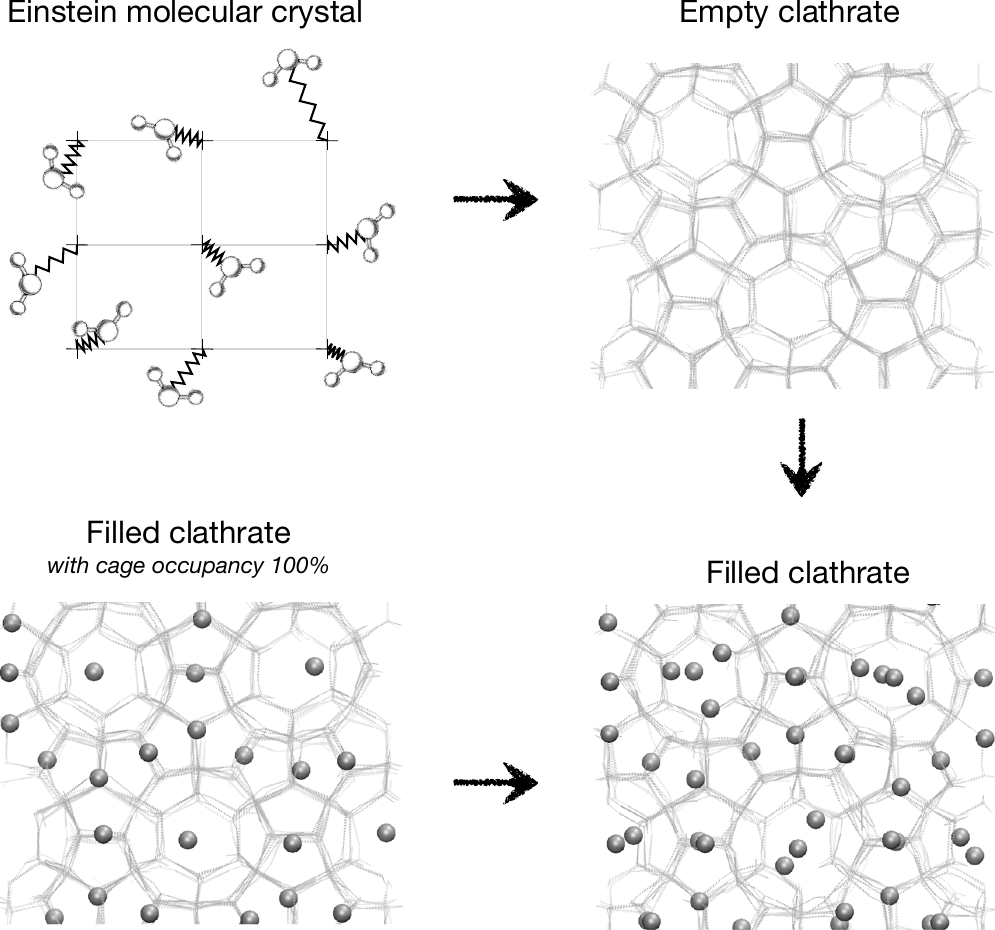}
    \caption{Two possible thermodynamic paths for computing the free energy of a filled clathrate. The first path starts from an Einstein crystal. The intermolecular interactions are then gradually switched on, and the springs that tether the molecules to lattice sites and specific molecular orientations are gradually switched off to obtain an empty clathrate, which is then filled with gas molecules. The second path starts with a filled clathrate where all cages contain exactly one molecule. This single-occupancy constraint is then lifted.}
    \label{fig:Fig1}
\end{figure}

In the present work, we study the relative stability of clathrate structures II and H in the case of pure methane and pure argon hydrates. For the Ar hydrate, a phase transition between sII and sH is known to occur experimentally at high pressure~\cite{Loveday2008,Hirai2023}. Such a solid-solid coexistence pressure cannot be determined via the direct coexistence method because the large free energy barrier between two phases prevents spontaneous phase transformations on simulation timescales. Consequently, explicit calculations of the free energy difference between the two crystalline phases are required to locate coexistence points. Such calculations can be performed by the Frenkel-Ladd (FL) method, the self-referential (SR) method \cite{Atamas2011,Atamas2015}, or by Lattice-Switch Monte Carlo\cite{Bruce2000} (LSMC).
The thermodynamic path employed in the FL method is illustrated in Fig.~\ref{fig:Fig1}.

 In both the SR and LSMC methods, the initial configuration can be either an empty clathrate or a fully occupied clathrate, with one molecule per cage. In this work, we deploy the LSMC method, starting from the fully occupied clathrate and also from the empty clathrate as a consistency check. The LSMC method provides an {\it a priori} shorter path than the FL method because it focuses on the free energy {\em difference} between two candidate structures rather than their absolute values. Additionally, it allows for a straightforward estimation of the uncertainty on the computed free energy difference and can be performed in a constant pressure ensemble, unlike the FL method, which requires constant volume conditions. 

Application of the LSMC method to gas hydrates is a novel step. Previous studies employing LSMC \cite{Underwood2014,Moro2024} predominantly focused on simpler, single-component systems —often atomic— where atoms were tightly bound to fixed lattice sites. This stands in contrast to structure H, where guest molecules can move freely within large cages. Additionally, the ability to evaluate the free energy difference between a singly-occupied clathrate and a clathrate with a cage occupancy determined by some gas chemical potential, is another key contribution. The entropy difference between these two states can be large, in particular for structure H, even when the gas chemical potential is such that all cages are occupied, on average, by one molecule.

Our objective is to assess the applicability of the Lattice-Switch Monte Carlo method to determine the free energy difference between two clathrate structures (at any pressure) and to determine a coexisting pressure between two clathrate structures. We consider the case of a gas hydrate in equilibrium with a gas reservoir (no excess water, all water molecules are part of the gas hydrate). We assume, for simplicity, a pure gas, but extending the methodology to a gas mixture is straightforward.  We assume, moreover, that the solubility of water in the gas phase is negligible. The chemical potential of the gas molecules depends then solely on pressure and temperature: $\mu_g = \mu_g(P, T)$.

In simulations, using a statistical ensemble that closely replicates experimental conditions is advantageous. The grand canonical ensemble, characterized by constant $N_w, \mu_g, V, T$—where $N_w$ is the number of water molecules and $\mu_g$ depends on both pressure ($P$) and temperature ($T$)—is commonly applied in gas hydrate absorption studies. However, its limitation lies in the necessity of a fixed volume. Since the average volume depends not only on pressure but also, to a lesser extent, on the amount of adsorbed gas at a given pressure, accurately determining the average volume for each pressure point can be a cumbersome task. This challenge can be circumvented by fixing the set of independent variables $N_w, \mu_g, P, T$. We refer to this as the $\Gamma$ ensemble~\cite{Wolf, Dubbeldam, Waage}. This ensemble is particularly well-suited for describing absorption in flexible frameworks like gas hydrates, as it allows the volume to adjust naturally in response to mechanical and gas pressures.

While absorption calculations are carried out in an open ensemble like the $\Gamma$ ensemble, the LSMC simulations, which yield the Gibbs free energy difference $\Delta G$ between two clathrate phases, must be conducted in a closed ensemble with a fixed number of gas molecules, for instance, $N_g = 0$ for an empty clathrate or $N_g = N_{\rm cages}$ for a fully occupied clathrate. In this work, we derive the free energy difference required to transition from the isothermal-isobaric ensemble (constant-($N_w, N_g, P, T$)) to the $\Gamma$ ensemble, in which $N_g$ is allowed to fluctuate. The LSMC simulation with $N_g = N_{\rm cages}$ is in fact conducted under the stricter condition that each cage is singly occupied. The free energy associated with relaxing this single-occupancy constraint is computed explicitly. The $\Gamma$ ensemble thus offers an elegant framework for determining the most stable clathrate structure at a given temperature and pressure when the hydrate is in contact with a gas reservoir: it is the phase with the lowest $\Gamma$ free energy (see Sect.~2).

By reframing the problem in terms of the $\Gamma$ free energy, one focuses directly on the free energy difference between the two competing phases. The difference, $\Delta\Gamma$, which depends on the pressure, conveniently describes the relative thermodynamic stability of two clathrate phases. This approach provides a more direct route to assessing relative stability as a function of pressure and locating coexistence than comparing convex-hull diagrams\cite{Hildebrandt1994,Mathews2026,Teeratchanan2015,berni2026,gondre2025} at multiple pressures. The calculations fully capture the energetic, volumetric, and entropic contributions to the free energy difference $\Delta \Gamma$. Fluctuations in volume and cage occupancy are naturally included. While these become negligible in the thermodynamic limit, treating them exactly is advantageous for finite systems.

The outline of this paper is as follows.  We begin in Sect.~\ref{sec:NwmugPT} by setting out the properties of the constant-$N_w,\mu_g,P,T$ ensemble from both thermodynamical and statistical mechanical perspectives. In Sect.~\ref{sec:Method}, we explain our simulation methodology, which utilizes a combination of LSMC computations, thermodynamic cycles, and thermodynamic integration. Sect.~\ref{Sct4} presents the model, numerical results, and their analysis. The last Sect.~\ref{sec:Ccl} summarises our principal findings.



\section{Properties of the $\Gamma$ ensemble {\large (constant $T,P,\mu_g,N_w$)}}
\label{sec:NwmugPT}

\subsection{Thermodynamic formulation}

In the $\Gamma$ ensemble, characterized by constant-$(T,P,\mu_g,N_w)$, the total energy $E$, volume $V$, and number of guest particles $N_g$ all fluctuate under the control of their respective conjugate fields. The conditions for equilibrium between two clathrate structures (phases) are equality of temperature ($T^{(1)} = T^{(2)}$), pressure ($P^{(1)} = P^{(2)}$) and of the chemical potentials of the two species of molecules (water: $\mu_w^{(1)} = \mu_w^{(2)}$ and gas: $\mu_g^{(1)} = \mu_g^{(2)}$) in both phases. Equal pressure means here equal mechanical stress in both phases. The conditions of equal temperature, pressure and chemical potential $\mu_g$ of the gas molecules are automatically satisfied when $(T,P,\mu_g,N_w)$ are fixed during a simulation run. (In the case of a gas mixture, one would have several gas chemical potentials instead of a single one.) Phase coexistence therefore occurs when $\mu_w^{(1)} = \mu_w^{(2)}$. Note that $\mu_w(T,P,\mu_g, N_w)$ is sensitive to gas content, and hence to the cage occupancies.

At constant $(T,P, \mu_g, N_w)$, the thermodynamic potential is given by the Legendre transform of the Gibbs free energy $G(T,P,N_g,N_w)=E-TS+PV=\mu_gN_g+\mu_wN_w$ with respect to $N_g$, namely
\begin{equation}
\label{def_gama}
    \Gamma(T,P, \mu_g,N_w) \equiv E - TS + PV - \mu_gN_g  = \mu_wN_w 
\end{equation}
The total differential of this potential is,
\begin{equation}
    d\Gamma = -S dT + V dP - N_g d\mu_g,
\end{equation}
which gives access to the average values of the conjugate variables, e.g.
\begin{eqnarray}
       V &=& \left.\frac{\partial \Gamma}{\partial P}\right|_{\mu_g,T}\label{eq:Vderiv}\\
       N_g &=& -\left.\frac{\partial \Gamma}{\partial \mu_g}\right|_{P,T}\label{eq:Ngderiv}
\end{eqnarray}

\subsection{Coexistence condition}
Since $\Gamma$ is minimised at equilibrium, phase coexistence at fixed $(T,P,\mu_g,N_w)$ occurs when
\begin{equation}
    \Gamma^{(1)} = \Gamma^{(2)}.
\end{equation}
With $\Gamma = N_w \mu_w$ and fixed $N_w$, this reduces to the usual condition $\mu_w^{(1)} = \mu_w^{(2)}$.

Having established that $\Gamma$ is a thermodynamic potential when expressed in terms of the independent variables $(T,P,\mu_g,N_w)$, the condition for phase coexistence reduces to $\Delta \Gamma = 0$ at fixed values of these variables. Although the fields $(T,P,\mu_g)$ are, in principle, independent, we assume that the pressure (stress) on the hydrate is entirely determined by the applied gas pressure. In this case, the gas chemical potential is no longer independent, but is fixed by the equation of state of the gas, $\mu_g = \mu_g(T,P)$. The coexistence condition is therefore given by a sign change of $\Delta \Gamma(P,T,\mu_g(T,P),N_w)$, i.e., the difference in thermodynamic potential evaluated along the constraint $\mu_g=\mu_g(T,P)$. This resolves the issue of comparing the stability of clathrate phases at fixed gas pressure while accounting for differences in stoichiometry: the appropriate quantity to compare is the $\Gamma$ potential, rather than the Gibbs free energy.

Our approach assumes that the total number of water molecules, $N_w$, is fixed. However, the number of water molecules per unit cell differs between clathrate structures: 46 in sI, 136 in sII, and 34 in sH. This mismatch is not problematic, since systems with identical $N_w$ can be constructed by considering different numbers of unit cells for each structure. For example, one sII unit cell (136 molecules) corresponds to four sH unit cells, while 23 sII unit cells (3128 molecules) correspond to 68 sI unit cells.
During a structural transition, the number of unit cells adjusts to accommodate the new lattice while preserving the total number of water molecules. As a result, $\Delta \Gamma$ can be consistently computed at fixed $N_w$.

\subsection{Statistical mechanical perspective}
\label{sec:GammaEns}

It is useful from the point of view of constructing a simulation scheme to develop the statistical mechanical formulation of the constant-$(T,P,\mu_g, N_w)$ ensemble to extend the thermodynamical description.  Within this ensemble, a microstate is fully characterised by:
\begin{enumerate}[topsep=3pt]
\setlength\itemsep{-0.2em}
    \item the number $N_g$ of guest particles
    \item the set of their position vectors $\{ {\bf r_i} \}^{N_g}$
    \item the set of position vectors $\{ {\bf r_j} \}^{N_w}$ of the fixed number of $N_w$ water particles
    \item the system volume $V$.
\end{enumerate}

\noindent
The probability of a microstate ${\color{blue}\ell}$ is
\begin{equation}
\label{eq:Pell}
P_\ell 
=
    P(\{ {\bf r}_i \}^{N_g}, \{ {\bf r}_j \}^{N_w}, V, N_g)
    =
    \frac{1}{Z_\Gamma}
\exp\!\left[-\beta\bigl(E_\ell + P V_\ell - N_{g,\ell}\mu_g\bigr)\right],
\end{equation}
where
\begin{multline}\label{eq:BoltzmanFactor}
Z_\Gamma(T,P,\mu_g,N_w)
=
\sum_\ell
\exp\!\left[-\beta\bigl(E_\ell + P V_\ell - N_{g,\ell}\mu_g\bigr)\right]
\\
=
\int_0^\infty dV
\sum_{N_g=0}^\infty
\left[
\prod_{i=1}^{N_g} \int d{\bf r}_i
\prod_{j=1}^{N_w} \int d{\bf r}_j
\right]
\exp\!\left[-\beta\bigl(E + PV - N_g\mu_g\bigr)\right],
\end{multline}
is the configurational partition function, $\beta=1/k_BT$, and the energy $E=E(\{ {\bf r_i} \}^{N_g}, \{ {\bf r_j} \}^{N_w})$ depends on the molecule positions. Using the Gibbs-Shannon formula, one can express the entropy as
\begin{align}
    S = -k_B \sum_\ell P_\ell \ln(P_\ell)
    &= -k_B \sum_\ell P_\ell  \left[
-\ln(Z_\Gamma) - \beta(E+PV-N_g\mu_g) \nonumber
    \right] \\[-1em]
    &=\, k_B \ln(Z_\Gamma) + \frac{1}{T}(\overline{E} + P \overline{V} - \overline{N_g} \mu_g),
\end{align}
which gives
\begin{alignat}{1} \notag
-k_BT \ln Z_\Gamma
&=
\overline{E} - TS + P \overline{V} - \overline{N_g} \mu_g \\
\label{eq:LinkGammaPartitionFct}
& \equiv \Gamma(T,P, \mu_g,N_w).
\end{alignat}
The free energy~\eqref{def_gama} is thus recovered. It is linked to the partition function~\eqref{eq:BoltzmanFactor} by the logarithmic relation~\eqref{eq:LinkGammaPartitionFct}, as expected. 

The statistical mechanics of the $\Gamma$ ensemble provides not only average values but also the full probability distributions of fluctuating quantities, such as the volume $V$, the number of guest particles $N_g$, the particle positions $\{{\bf r}_i\}^{N_g}, \:\{{\bf r}_j\}^{N_w}$, and any derived observable. These fluctuations are directly accessible in Monte Carlo simulations.

At fixed $(T,P,\mu_g,N_w)$, the probability of observing a given energy $E$ is obtained by summing Eq.~\eqref{eq:Pell} over all microstates with that energy. Introducing the constrained partition sum
\begin{equation}
Z_\Gamma(E) = \sum_{\ell:\,E_\ell=E} \exp\!\left[-\beta\bigl(E + PV_\ell - N_{g,\ell}\mu_g\bigr)\right],
\end{equation}
one finds
\begin{equation}\label{eq:ProbMacrostate1}
P(E \mid T,P,\mu_g,N_w)
= \frac{Z_\Gamma(E)}{Z_\Gamma}.
\end{equation}
Defining the constrained free energy
\begin{equation}
\Gamma(E) = -k_B T \ln Z_\Gamma(E),
\end{equation}
this can be written in the compact form
\begin{equation}
P(E) = \exp\!\bigl[-\beta\bigl(\Gamma(E)-\Gamma\bigr)\bigr],
\end{equation}
where $\Gamma = -k_B T \ln Z_\Gamma$ is the thermodynamic potential of the $\Gamma$ ensemble [Eq.~\eqref{def_gama}]. Here $\Gamma(E)$ incorporates both the density of states and the fluctuations of $V$ and $N_g$ at fixed energy $E$.

The thermodynamic relations \eqref{eq:Vderiv} and \eqref{eq:Ngderiv} express the average volume and number of gas particles. It is instructive to make use of the full probability distribution~\ref{eq:Pell} to derive the average volume and average number of gas molecules. Starting from eq.~\eqref{eq:LinkGammaPartitionFct}, one has
\begin{equation}
\begin{split}
\frac{\partial \Gamma}{\partial P}&=-k_BT\frac{1}{Z_\Gamma}\frac{\partial Z_\Gamma}{\partial P}\nonumber\\[-0.8em]
&=
    -k_BT\frac{1}{Z_\Gamma}\int_0^\infty dV \prod_{i=1}^{N_g}\int d{\bf r_i} \prod_{j=1}^{N_w} \int \ d{\bf r_j} \\[-0.9em]
    &\qquad\times\sum_{N_g=0}^\infty (-\beta V) \exp{\left(-\beta[E+PV-N_g\mu_g]\right)}
\nonumber\\[-0.8em]
&= \int_0^\infty V p(V)dV
\label{eq:link_meanV_DeltaGamma}
\;\equiv\;  \langle V\rangle\:,
\end{split}
\end{equation}
which is equivalent to \ref{eq:Vderiv} and
\begin{equation}
\begin{split}
\frac{\partial \Gamma}{\partial \mu_g}&=-k_BT\frac{1}{Z_\Gamma}\frac{\partial Z_\Gamma}{\partial \mu_g}\nonumber\\[-0.8em]
&=
-k_BT\frac{1}{Z_\Gamma}\int_0^\infty dV \prod_{i=1}^{N_g}\int d{\bf r_i} \prod_{j=1}^{N_w} \int \ d{\bf r_j} \\[-0.9em]
&\qquad\times\sum_{N_g=0}^\infty (\beta N_g) \exp{\left(-\beta[E+PV-N_g\mu_g]\right)}
\nonumber\\[-0.8em]
&= -\sum_{N_g=0}^\infty N_g p(N_g) 
\label{eq:link_meanNg_DeltaGamma}
\;\equiv\; -\langle N_g\rangle\:,
\end{split}
\end{equation}
which is equivalent to \ref{eq:Ngderiv}. The fluctuations (variance) of $V$ and $N_g$ in the $\Gamma$ ensemble are related to the second derivatives of $\Gamma$ with respect to $P$ or to $\mu_g$.
In addition to describing the fluctuations, the statistical mechanical formulation will prove crucial for deriving a formula [see later eq.~\eqref{eq:FluctCorr}] for the free energy difference between a gas hydrate in which all cages contain one guest molecule (fully occupied clathrate) and a gas hydrate with the same gas content but without such constraint on the location of the guest molecules.

\section{Thermodynamic cycles for determining $\Delta\Gamma$ and Lattice-Switch Monte Carlo}
\label{sec:Method}

We now present the thermodynamic cycles used to compute the free energy difference $\Delta \Gamma$ between two clathrate structures and to determine the pressure at which $\Delta \Gamma = 0$ (coexistence). The starting point is the Gibbs free energy difference $\Delta G = \Delta U + P \Delta V - T \Delta S$ between the two structures, which can be obtained using any suitable free-energy method. In this work, $\Delta G$ is computed using lattice-switch Monte Carlo (LSMC), briefly outlined in Sect.~\ref{subsec:LSMC}.

LSMC allows one to work directly with either empty clathrates (Sect.~\ref{subsec:CycleEmpty}), or fully filled ones (Sect.~\ref{subsec:CycleFilled}), thereby avoiding integration to reference states such as Einstein crystals. However, it requires that the numbers of water and guest molecules ($N_w$ and $N_g$) are identical in both phases.  To obtain equilibrium free energy differences at a given $(P,T)$, for which the values of $N_g$ in the two phases generally differ from those in the empty or fully filled state, and from each other,  one needs to relax the constraint of fixed $N_g$ by transforming to the $\Gamma$ ensemble. Doing so allows $N_g$ to fluctuate under the control of $\mu_g=\mu_g(P,T)$. This transformation is trivial in the case of a cycle based on LSMC between empty clathrates, because $\Delta G=\Delta\Gamma$, but requires additional calculation in the case of a filled clathrate. 

Once in the $\Gamma$ ensemble, and with knowledge of $\Delta G$ from the LSMC step, equilibrium free energy differences and coexistence conditions at a given temperature (Sect.~\ref{subsec:coexistence}) are obtained via thermodynamic integration with respect to $P$ and $\mu_g$. 

\subsection{Cycle starting from empty clathrates}
\label{subsec:CycleEmpty}

\begin{figure}[h]
\centering
  \includegraphics[width=0.9\linewidth]{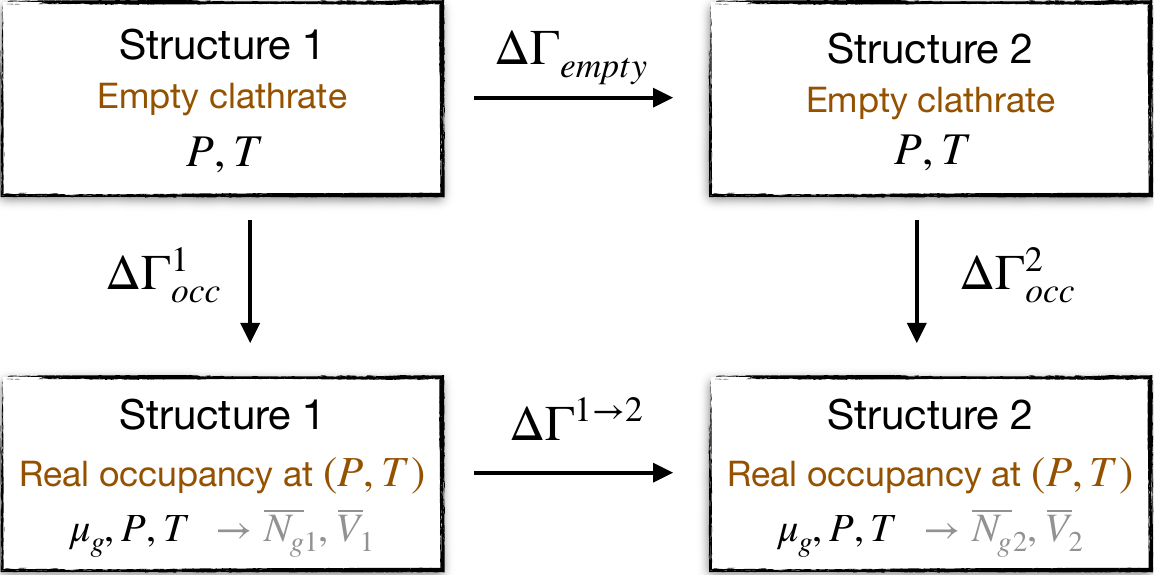}
  \caption{Thermodynamic cycle starting from two empty clathrates. These clathrates are then filled with gas molecules by performing successive Monte-Carlo simulations in the $\Gamma$ ensemble along the path for $\mu_g$ from $-\infty$ to $\mu_g(P,T)$. The free energy of filling is given by eq.~\eqref{eq:DeltaGammaOcc}. In the filled state, the gas pressure is identical to the mechanical stress $P$.
  }
  \label{fig:cycle_simple}
\end{figure}

The Gibbs free energy $G = N_w \mu_w + N_g \mu_g$ of an empty ($N_g = 0$) clathrate reduces to the free energy $\Gamma = N_w \mu_w$. Once the free energy difference $\Delta G_{\text{empty}}=\Delta \Gamma_{\text{empty}}$ between two empty clathrate structures is known at some temperature $T$ and pressure $P$ (see Sect.~\ref{subsec:LSMC}), one can compute the corresponding $\Delta \Gamma$ between the two filled clathrates (where the filling depends on the chosen $P$ and $T$) by thermodynamic integration along an absorption isotherm.\cite{Wierzchowski,Jensen2010} The corresponding thermodynamic cycle is shown in Fig.~\ref{fig:cycle_simple}. The free energy difference between the filled clathrate structures 1 and 2 at real occupancy is denoted $\Delta\Gamma^{1\rightarrow 2}$.

$\Delta\Gamma_{\text{occ}}^\alpha$ is the free energy difference 
 between a filled and an empty clathrate in structure $\alpha$ at the same temperature and pressure. In the empty clathrate, the gas chemical potential is minus infinity, while it is 
 $\mu_g(P,T)$ in the filled clathrate. $\Delta\Gamma_{\text{occ}}$ can therefore be computed with 
 \begin{equation}\label{eq:DeltaGammaOcc}
\Delta \Gamma_{\text{occ}}=\int_{-\infty}^{\mu_{g}} \frac{\partial \Gamma}{\partial \mu_g
}d\mu_g = \int_{-\infty}^{\mu_{g}}  \langle N_g \rangle  
\,d\mu_g
\end{equation}
where we have used Eq.~\eqref{eq:Ngderiv} and, for simplicity, suppressed the superscript $\alpha$. This corresponds to integrating $\langle N \rangle(\mu_g)$ along an absorption isotherm at fixed $T$ and mechanical stress $P$. The upper limit $\mu_g(P,T)$ is obtained from the equation of state of the gas.

\subsection{Cycle starting from filled clathrates}
\label{subsec:CycleFilled}

\begin{figure}[h]
\centering
  \includegraphics[width=0.9\linewidth]{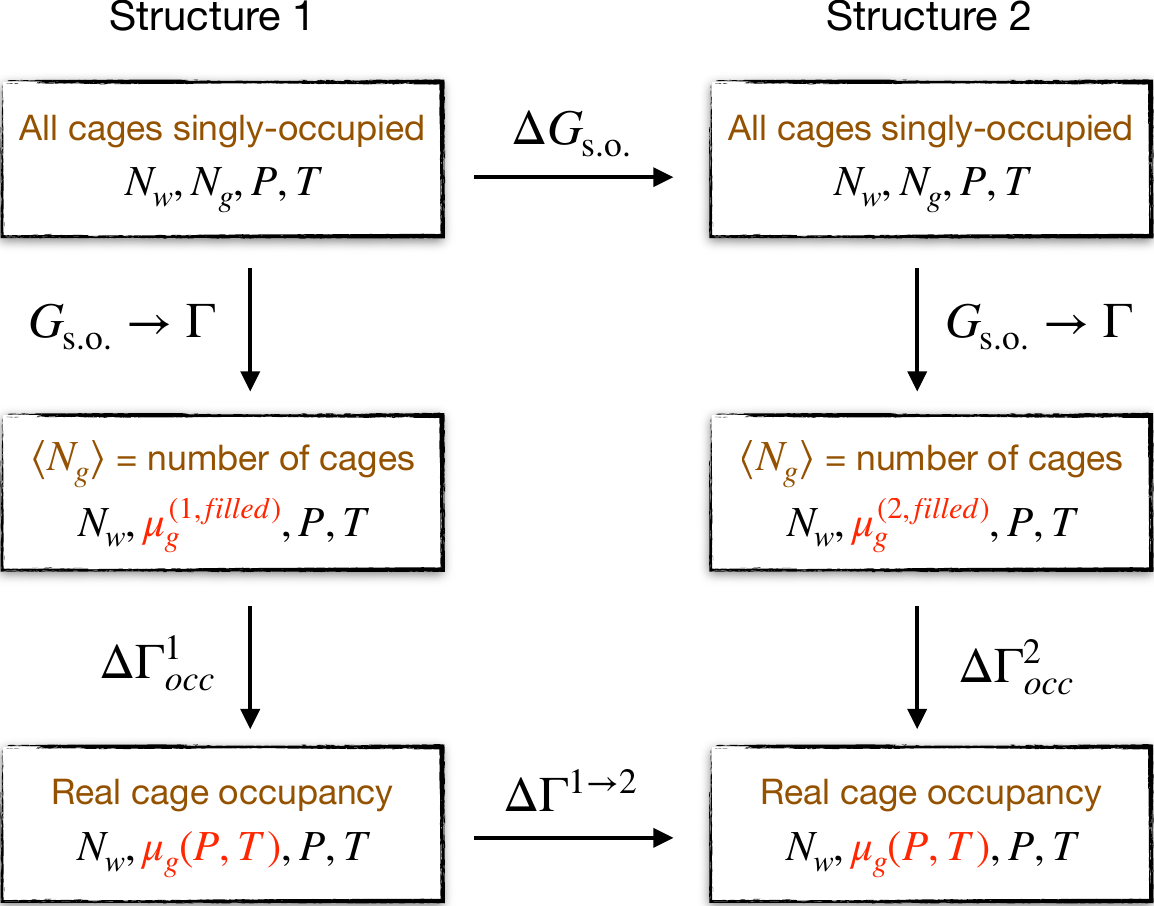}
\caption{Cycle to compute $\Delta\Gamma^{1\to 2}$ between two clathrate structures
starting from the Gibbs free energy difference between the two structures with all cages singly occupied. One first changes the ensemble by allowing $N_g$ to fluctuate around $N_{\text{cages}}$ and lifting the constraint that all cages are singly occupied. The corresponding free energy change is given by eq.~\eqref{eq:GsoToGamma}. The cage occupancy is then adjusted to correspond to the applied pressure $P$.}
  \label{fig:cycle_filled}
\end{figure}

The starting reference point is now the filled clathrate structures, where each cage is occupied by one gas molecule. Then, obviously, $N_g = N_{\text{cages}}$.
Once the Gibbs free energy difference $\Delta G_{\text{s.o.}}$ between two such structures is known (s.o. stands for `single occupancy'), one can use the thermodynamic cycle in Fig.~\ref{fig:cycle_filled} to compute the free energy difference $\Delta\Gamma^{1 \to 2}$ between the two structures with the correct cage filling at the considered $P$ and $T$.

Since $\Gamma = G - \mu_g N_g$ [recall eq.~\eqref{def_gama}], one needs to subtract the number of gas molecules times their chemical potential. It is important to realize that (a) the chemical potential of the guest molecules is not identical in both structures when they are fully occupied because their environment is not the same, and (b) the Gibbs free energy $\Delta G_{\text{s.o.}}$ is computed in an ensemble where $N_g$ cannot fluctuate and under the constraint that all cages are singly occupied. By performing Monte Carlo simulations in the $\Gamma$ ensemble, one can determine the chemical potential $\mu_g^{\alpha,\, \text{filled}}$ at which structure $\alpha$ is filled in the sense $\langle N_g \rangle = N_{\text{cages}}$. Then, the switch from $G$ to $\Gamma$ involves
\begin{equation}
\label{eq:EnsembleSwitch_Legendre}
     \Delta G_{\text{s.o.}} - (\mu_g^{\alpha=2,\, \text{filled}} - \mu_g^{\alpha=1,\, \text{filled}}) N_{\text{cages}}.
\end{equation}
However, as we now explain, this formula alone is insufficient to describe the ensemble change because it neglects fluctuations in the value of $N_g$ and, more importantly, the removal of the constraint requiring single occupancy of all cages. 

\subsubsection*{Accounting for fluctuations and non-singly occupied cages}

Let us define $\eta=N_{\text{single}}/N_{\text{cages}}$ as the fraction of singly occupied cages and $\Gamma(\eta=1)$ as the free energy of the singly-occupied states in the $\Gamma$ ensemble which match to the constant $N_w, N_g, P, T$ LSMC simulations. The constrained free energy 
\begin{equation}
    \Gamma(\eta) = -k_{\rm B}T\, \ln Z_\Gamma(\eta)
\end{equation}
involves $Z_\Gamma(\eta)$ which is given by eq.~\eqref{eq:BoltzmanFactor} but where only states with a given value of $\eta$ are considered. The probability that a fraction $\eta$ of the cages are singly-occupied is 
\begin{equation}
\label{eq:prob-eta}
    P(\eta) = e^{-\beta (\Gamma(\eta) - \Gamma)}
\end{equation}
where $\Gamma$ is the free energy at gas chemical potential $\mu_g^{\alpha,\,\text{filled}}$, {\it i.e.\@} when the number of gas molecules fluctuates around the average $\langle N_g \rangle = N_{\text{cages}}$. Introducing the constraint $\eta = 1$ increases the free energy by
\begin{equation}
    \label{eq:def DeltaGammaSO}
    \Delta \Gamma^\alpha_{\text{s.o.}}
    = \Gamma(\eta = 1) - \Gamma > 0.
\end{equation}
Combining eqs.~\eqref{eq:prob-eta} and \eqref{eq:def DeltaGammaSO}, we find
\begin{equation}
    \label{eq:FluctCorr}
    \Delta \Gamma^\alpha_{\text{s.o.}} = -k_{\rm B}T \, \ln(P(\eta=1)).
\end{equation}
Suppressing the single-occupancy constraint thus lowers the free energy by $\Delta \Gamma^\alpha_{\text{s.o.}}$, which is directly related to the probability of observing states with all cages singly-occupied in the $\Gamma$ ensemble at $\mu_g = \mu_g^{\alpha,\,\text{filled}}$. In other words, $-\Delta \Gamma^\alpha_{\text{s.o.}}$ is the requisite free energy difference that accounts for fluctuations in both $N_g$ and cage occupancy\footnote{Notice that transforming free energies with a Legendre transform, as in eq.~\eqref{eq:EnsembleSwitch_Legendre}, holds only in the thermodynamic limit because it neglects fluctuations. Indeed, the present calculation implies that allowing a variable (here a number $N$ of molecules) to fluctuate introduces a free energy shift $-k_{\rm B}T \ln(P(N))$, where $P(N)$ is the probability of observing the (previously fixed) value $N$. This shift is a fluctuation correction that allows switching between ensembles exactly in finite-size systems. It can be disregarded only in the thermodynamic limit. $\Delta\Gamma_{\text{s.o.}}^\alpha$ includes not only such a fluctuation correction but also a correction for lifting the single-occupancy constraint. These two contributions emerge clearly when expressing the probability $P(\eta=1)$ as the product $P(N_{\text{single}} = N_{\text{cages}} |  N_g=N_{\text{cages}}) P(N_g=N_{\text{cages}})$. The second factor provides the fluctuation correction to $\Delta\Gamma_{\text{s.o.}}^\alpha$, 
which is independent of system size, while the first factor with the conditional probability provides the single-occupancy correction, whose contribution to $\Delta\Gamma_{\text{s.o.}}^\alpha$ is proportional to the system size.
} \\

The step $G_{\text{s.o.}} \to \Gamma$ in the cycle is performed therefore with
\begin{equation}
\label{eq:GsoToGamma}
\Delta(G_{\text{s.o.}} \to \Gamma) =
 - (\mu_g^{\alpha=2,\, \text{filled}} - \mu_g^{\alpha=1,\, \text{filled}}) N_{\text{cages}} 
 - (\Delta \Gamma^{\alpha=2}_{\text{s.o.}}
 - \Delta \Gamma^{\alpha=1}_{\text{s.o.}}).
\end{equation}
The probability $P(\eta)$ in~\eqref{eq:FluctCorr} can be measured in a Monte Carlo simulation. If this probability is very low because a cage is likely to be occupied by more than one molecule (e.g. in structure H or at high pressures), the sampling of the configuration space needs to be biased to ensure that the states with $\eta = 1$ are sampled sufficiently often in the course of the simulation. This can be done by using free energy sampling methods like transition-matrix Monte Carlo\cite{Fitzgerald1999} or Wang-Landau\cite{Wang2001}.

\noindent We note in passing that one can establish (see Appendix~\ref{AppendixA}) an upper bound for \(\Delta\Gamma^\alpha_{\text{s.o.}}\) under the assumption that interactions between guest molecules increase the probability of single occupancy compared to the ideal case, where guest molecules are non-interacting. This upper bound makes clear the potentially large entropic contribution in \(\Delta\Gamma^\alpha_{\text{s.o.}}\).\\

The second step in the cycle involves the free energy change $\Delta\Gamma_{\text{occ}}^\alpha$ as in the first cycle (Fig.~\ref{fig:cycle_simple}) but where the lower bound in  eq.~\eqref{eq:DeltaGammaOcc} for $\Delta\Gamma_{\text{occ}}^\alpha$ is replaced by $\mu_g^{\alpha,\, \text{filled}}$ since the starting point is now the filled clathrate.

\subsection{Locating coexistence}
\label{subsec:coexistence}

To determine the coexistence pressure $P_{\text{coex}}$ for a given temperature, we vary $P$ to find the value for which  $\Delta \Gamma^{1\to 2}(P_{\text{coex}}) = 0$. When the pressure changes, the cage occupancy changes as well. One can generalise eq.~\eqref{eq:DeltaGammaOcc} to compute the free energy difference $\Delta\Gamma$ associated with a change of pressure (from $P$ to $P'$) and of cage occupancy (from $\mu_g$ to $\mu'_g$). Specifically, we want to compute $\Delta \Gamma$ between a clathrate in state ($N_w, \mu_g, T, P$) (here $\mu_g$ can have any value, e.g. $-\infty$ for an empty clathrate) and a filled clathrate at another pressure $P'$ with the corresponding cage filling (state $N_w, \,\mu_g'=\mu_g(T,P'),\, T,\, P'$). It suffices to integrate along any path from $(\mu_g, P)$ to $(\mu_g', P')$. From the chain rule
\begin{align}\label{eq:DeltaGamma}
\Delta \Gamma_{P\to P'}
&=\int_{P}^{P'} \left(\frac{\partial \Gamma}{\partial P
}\right)dP + \int_{\mu_{g}}^{\mu_{g}'} \left(\frac{\partial \Gamma}{\partial \mu_g
}\right)d\mu_g \\
&= \int_{P}^{P'}\langle V\rangle dP - \int_{\mu_{g}}^{\mu_{g}'} \langle N_g\rangle d\mu_g.
\label{eq:DeltaGammaUse}
\end{align}
Since $\Gamma$ is a state function, the result is path independent, allowing $P$ and $\mu_g$ to be varied simultaneously. If, at the starting point, the gas chemical potential satisfies $\mu_g=\mu_g(T,P)$ (i.e.\ the gas pressure matches the mechanical stress), it is natural to maintain this condition along the integration path. The path can then be parametrized as $(\tilde{P}, \mu_g(T,\tilde{P}))$, with $\tilde{P}$ varying from $P$ to $P'$.

\begin{figure}[h]
    \centering
    \includegraphics[width=0.9\linewidth]{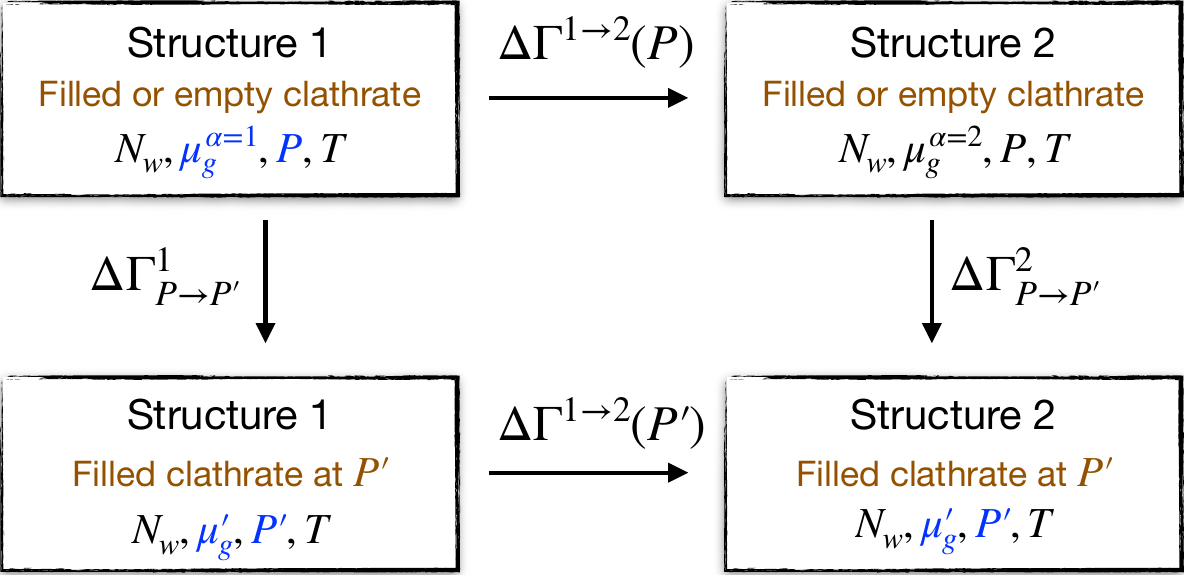}
    \caption{Cycle to compute $\Delta\Gamma^{1\to 2}$ at any pressure $P'$ starting from $\Delta\Gamma^{1\to 2}$ at a reference pressure $P$. This cycle is used to search for a coexistence pressure where $\Delta \Gamma^{1\to 2}(P') = 0$. The free energy change $\Delta \Gamma_{P \to P'}$ is given by eq.~\eqref{eq:DeltaGammaUse} and can be computed from multiple Monte-Carlo simulations in the $\Gamma$ ensemble.}
    \label{fig:cycle_coex}
\end{figure}

The corresponding thermodynamic cycle used to compute $\Delta \Gamma^{1 \to 2}$ at pressure $P'$ is shown in Fig.~\ref{fig:cycle_coex}. The evaluation of $\Delta \Gamma_{P \to P'}$ requires measuring $\langle V \rangle$ and $\langle N_g \rangle$ in a series of $\Gamma$-ensemble simulations at different pressures and gas chemical potentials.

\subsection{Lattice-Switch Monte Carlo}
\label{subsec:LSMC}

Lattice-Switch Monte-Carlo\cite{Bruce1997,Bruce2000,Jackson2002,Bruce2003,Jacksonthesis} is a powerful simulation method that permits the direct measurement of the Gibbs free energy difference $\Delta G$ for crystalline molecular systems. It circumvents the need to calculate absolute free energies by thermodynamic integration to reference states and the difficulty of accurately estimating a small free energy difference by subtracting two large numbers. The method directly provides a fundamental ingredient of the thermodynamic cycles introduced in the previous section, namely the Gibbs free energy difference between two structures at a given pressure and temperature.

LSMC exploits a mapping between two distinct crystalline phases having equal numbers of molecules. Each phase is described by a set of reference lattice sites, which are the ideal molecular positions in the perfect crystal; each molecule’s instantaneous position and orientation are expressed as a displacement and rotation relative to its associated site. This mapping is incorporated into a Monte Carlo move, the lattice-switch move, which instantly transforms one crystalline structure into another while preserving the relative displacements and orientations, and replacing the underlying reference lattice. This provides a reversible path that connects the two phases of interest, giving access to the free energy difference $\Delta G$ between them. LSMC can be used for fluids \cite{wilding_freezing_2000,McNeil-Watson2006} but it is especially efficient for solid-solid transitions. It computes the difference $\Delta G$ (rather than absolute free energies) by collecting statistics along the reversible path between the two phases. Several lattice-switch transitions between the two structures must occur during the course of an LSMC simulation to accurately estimate $\Delta G$. The ``reaction'' (advancement) coordinate along the reversible path is the lattice-switch order parameter~$M(\sigma)$, which is a function of the energy difference $E_\sigma - E_{\sigma'}$ with $E_\sigma$ the energy of the system in microstate $\sigma$ and $E_{\sigma'}$ the energy of the corresponding state after a lattice-switch move. 
The standard choice for this function is
\begin{equation}
\label{eq:M(sigma)}
    M(\sigma) = 
    \begin{cases}
        E_\sigma - {E_\sigma'} & \text{if microstate $\sigma$ in phase 1}\\
        |E_\sigma - {E_\sigma'}| & \text{if microstate $\sigma$ in phase 2}\\
    \end{cases}
\end{equation}
Typical configurations belonging to phase 1 and phase 2, are characterized by large negative and positive values of $M$, respectively. According to the Metropolis acceptance criterion, a lattice-switch move is likely to be accepted only when $M$ is small ($\,\lesssim k_B T$).
Configurations satisfying this condition are referred to as  {\em gateway} states since they allow transitions between the two crystalline structures. In practice, the unbiased Monte Carlo sampling visits these states only rarely, if at all. In order to reach them, we deployed Transition-Matrix Monte Carlo (TMMC) \cite{Smith1996,McNeil-Watson2006} to compute the requisite bias (weight) function. TMMC is very efficient for overcoming steep free energy barriers, such as the barrier to reaching the gateway states, which can be quite high if the two phases differ significantly. 

The LSMC simulation calculates the free energy difference between the phases from the collected statistics via
\begin{equation}\label{eq:dG}
   \Delta G = -k_{B}T\ln \frac{P_{2}}{P_{1}} 
\end{equation}
Here
$P_{1}$ and $P_{2}$ are the equilibrium probabilities for the system to be found in microstates typical of phases 1 and 2, respectively. In practice, these two probabilities are obtained by computing the (unbiased) integrated probability for the system to be found in configurations with order parameter $M<0$, which correspond predominantly to phase 1, and with $M>0$, which correspond predominantly to phase 2.  The bias introduced during the simulation to enhance sampling of gateway states is removed a posteriori when reconstructing the probability distribution of the order parameter. Further details of the LSMC and TMMC methods, including the construction of the biasing weights and the procedure for unbiasing the sampled distributions, as well as the numerical implementation used here, can be found in OM's thesis .~\cite{Moro2026}


\newpage
\section{Numerical results}
\label{Sct4}
\subsection{Model and parameters}\label{sec:model}

We consider the transition between the clathrate structures II and H in methane and argon hydrates. The calculations have been performed using the open-source multipurpose MC simulation program DL\_MONTE \cite{dlmonte,dlmontesource}. Additional explanations on the use of LSMC and TMMC methods within the context of DL\_MONTE can be found in the manual and tutorials.~\cite{dlmontemanual,dlmontetutorial}

Our orthorhombic simulation box contains $N_{w}=272$ water molecules forming either 2 unit cells of sII, or 4 unit cells of sH. In both structures, the total number of cages is 48. The positions and orientations of the water molecules in sII and sH were taken from Ref.~\cite{Takeuchi2013} where the lattice parameters (at $T=0$) are given as :
\begin{itemize}[noitemsep,topsep=0pt]
    \item for sII : $a = 17.31$ \AA
    \item for sH : $a=12.21$ \AA, $b=21.15$ \AA, $c=10.14$ \AA.
\end{itemize}
The unit cell of sII (containing 136 H$_2$O mols) was duplicated along the $x$-axis (2$\times$1$\times$1) while the one for sH (containing 68 H$_2$O mols) was duplicated along the $x$ and $z$ axes (2$\times$1$\times$2).

Our clathrate model employs the TIP4P/Ice water model \cite{WaterModel}. In this model, the Lennard-Jones parameters associated with the oxygen atom are $\epsilon=0.88216$\,kJ/mol and $\sigma=3.1668$\,\AA. Methane is described using the united-atom model of the TraPPE forcefield\cite{trappe}: $\epsilon=1.23054$ kJ/mol and $\sigma=3.73$\,\AA. For argon, we used the parameters  $\epsilon=0.9960$ kJ/mol and $\sigma=3.408$\,\AA\ proposed in Ref.~\cite{vdW1959}.  The interaction between different molecular species is given by the Lorentz-Berthelot mixing rule. Electrostatic interactions were computed using the Ewald method. The Lennard-Jones interactions were truncated at $r_{\text{cut}} \approx 8.7$ \AA\, and the standard long-range correction was included.

The chemical potential $\mu_{g} = \mu(P,T)$ in the guest (CH$_4$, Ar) phase was computed using the Lennard-Jones equation of state of Ref.~\cite{Thol2016}.  The fugacity coefficient predicted by this equation agrees well with experiments over a wide range of pressures.\cite{Moro2026}
The LSMC simulations were performed at a pressure of 30 bar and a temperature of 294~K. The number $N_g$ of gas molecules must be identical in both clathrate structures for the lattice-switch move to be well defined. We have considered two limiting cases: empty clathrates ($N_g=0$), and fully-occupied clathrates ($N_g=N_\text{cages} = 48$), in which all cages contain one molecule. 
The mapping of the lattice-switch move between lattice sites in sII and sH was chosen arbitrarily by following the order of the molecules within the unit cells given in Ref.~\cite{Takeuchi2013} (see also Sect.~\ref{sct:correspondence}).

\subsection{Free energy difference $\Delta G^{\text{II}\to\text{H}}$ for empty clathrates}
\label{sec:LSMCResEmpty} \label{sec:ResEmpty}

The free energy difference between empty clathrates is the quantity required for the first stage of the thermodynamic cycle in Fig.~\ref{fig:cycle_simple}. We computed $\Delta G^{\text{II}\to\text{H}}_{\text{empty}} = \Delta \Gamma_{\text{empty}}^{\text{II}\to\text{H}} = N_w \Delta \mu_w$ (because $N_g = 0$) with LSMC at $30$ bar and $294$ K.

\begin{figure}[htb]
 \centering
 \begin{subfigure}[t]{0.45\textwidth}
     \centering
     \includegraphics[width=\linewidth]{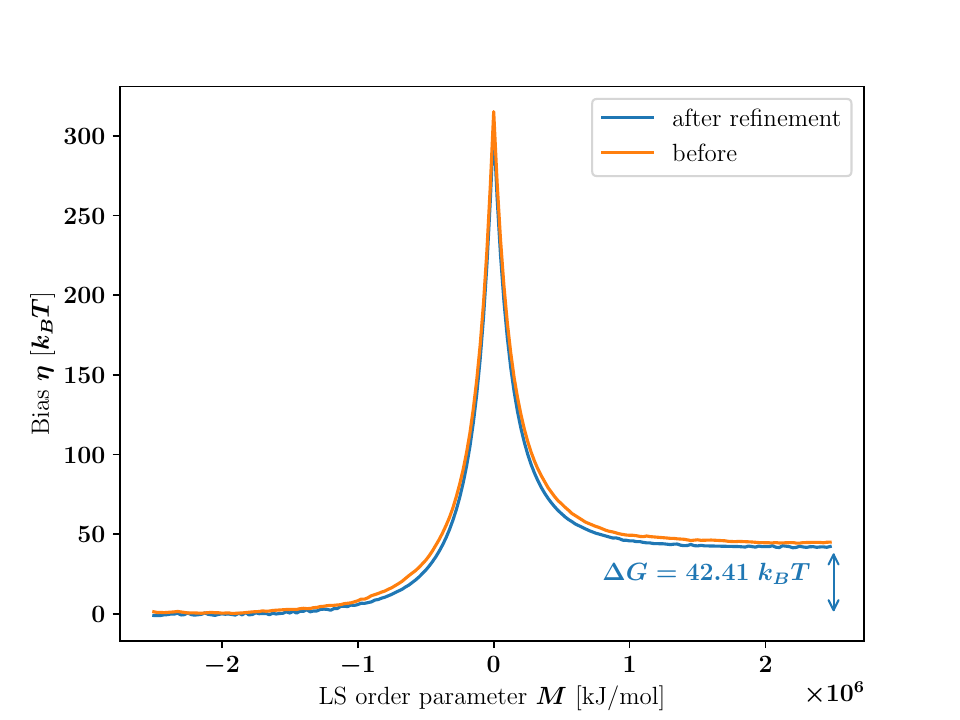}
 \end{subfigure}
 \begin{subfigure}[t]{0.45\textwidth}
    \centering
    \includegraphics[width=\linewidth]{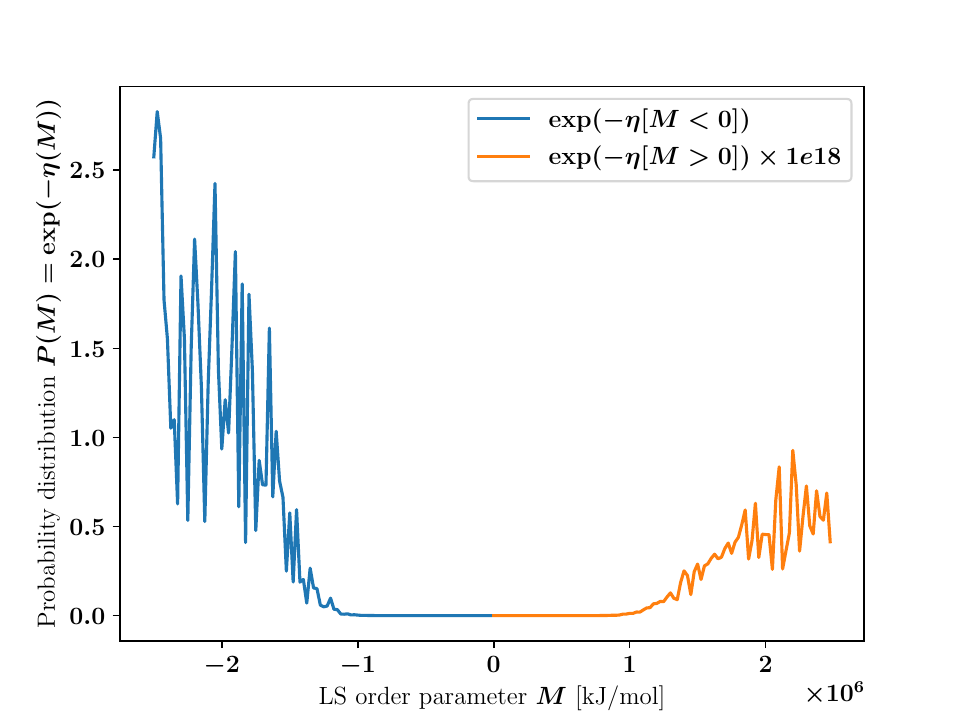}
 \end{subfigure}
     \caption{(Top) Free energy profile (bias function $\eta$) along the path between structures II and H parametrized by the lattice-switch order parameter~$M$ [eq.~\eqref{eq:M(sigma)}] at $30$ bar and $294$ K. The refined bias function was obtained by applying eq.~\eqref{etaCorr}; (Bottom) Corresponding probability distribution $P(M)=\exp(-\eta(M))$ for the refined bias function, after further smoothing using data from LSMC production simulations. The curve is multiplied by $10^{18}$ in region $M \geq 0$ to make the peak associated with structure~H visible.}
 \label{fgr:Empty_LSMC}
\end{figure}

The first step is to measure typical values of the lattice-switch order parameter (o.p.), eq.~\eqref{eq:M(sigma)}, in each phase to determine the range of order parameter values along the path connecting phase 1 (sII) to phase~2 (sH). Short single-phase $NPT$ simulations showed that the range $[-2.5,\,2.5]\times 10^6$ kJ/mol begins near the most probable o.p. value in sII and ends near the most probable value in sH (see Fig.~\ref{fgr:Empty_LSMC}, right plot). Restricting the sampling to this range focuses the computational effort on the most challenging part of the calculation, namely the accurate determination of the free-energy barrier separating the two phases. The contributions to $\Delta G^{\text{II}\to\text{H}}$ from states outside this range, i.e., from the neglected halves of the probability peaks in each phase, are added \textit{a posteriori}. 

The second step is to compute the free energy profile along the path. This profile is subsequently used as the bias function in the production LSMC simulation. The free energy profile was computed using the TMMC method. We discretised the o.p. range into \SI{20000}{} bins, a number chosen so that the difference of the bias function between adjacent bins does not exceed approximately $1.5 k_BT$. 

To construct the bias function, we employed 20 replicas (processors), each running for \SI{5.44e7}{} MC steps in a different window of the range. The resulting estimate of the bias function is labeled "before refinement" in Fig.~\ref{fgr:Empty_LSMC}. The two expected minima of the bias function $\eta(M)$, corresponding to peaks of the probability distribution $\exp(-\eta)$ associated respectively with sII and sH (Fig.~\ref{fgr:Empty_LSMC}(bottom)), are barely visible on this plot. This is mainly a consequence of the scale of the vertical axis, which spans the large free energy barrier ($\approx 300\,k_{B}T$) separating the two phases. From this preliminary profile, we obtain a first estimate of the free energy difference between the empty clathrate structures II and H, of approximately $42.4 k_B T$. 

For the production LSMC simulations, the bias function is fixed and must be sufficiently accurate to achieve approximately uniform sampling across the full range of the order parameter. This allows gateway states, where lattice-switch transitions can occur, to be visited frequently enough, while preventing the simulation from becoming trapped in particular regions of the order-parameter space.
 
We performed 20 such simulations, each lasting for \SI{5.44e8}{} MC steps, starting from a gateway state. These simulations sampled the central region well (corresponding to $M \in [-0.5,0.5]\times10^6$ kJ/mol), but only rarely reached the typical states of the two phases located at $|M|\simeq \SI{2e6}{}$ kJ/mol. This behaviour indicated that the bias function required improvement near $|M|\simeq \SI{0.5e6}{}$.

Using the data produced by these twenty production simulations with the applied bias $\eta(M)$, we refined the bias function according to\footnote{This expression follows from the definition of the optimal bias function, $\eta_{\mathrm{optimal}}(M)=-k_B T \ln P(M)$, which yields flat sampling, together with the biased distribution $P_\eta(M)=P(M)\,e^{\eta(M)}$ obtained when a bias $\eta(M)$ is applied. If $P_\eta(M)$ were known with infinite precision, the corrected bias $\eta_{\mathrm{corr}}$ would coincide with the optimal bias $\eta_{\mathrm{optimal}}$.}

\begin{equation}
\label{etaCorr}
    \eta_{\text{corr}}(M) = \eta(M) -\ln h_\eta(M)
\end{equation}
where $h_\eta(M)$ is the visited-states histogram of the production simulations. Eq.~\eqref{etaCorr} corrects the bias function such as to produce a flat histogram, $\eta_{\text{corr}} = \eta$. The refined bias function is shown in Fig~\ref{fgr:Empty_LSMC} (curve labeled ``after refinement''). 

\begin{figure}[htb]
 \centering
     \includegraphics[width=0.75\linewidth]{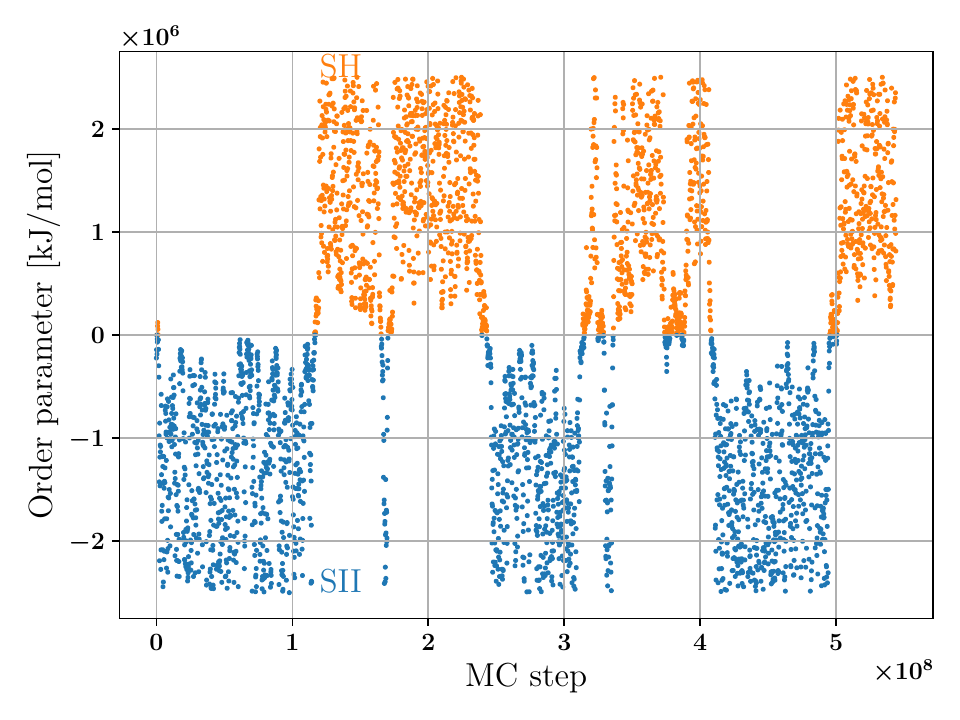}
\caption{
Lattice-switch order parameter, Eq.~\eqref{eq:M(sigma)}, during a representative LSMC simulation (one of 20 runs) for empty clathrates, showing multiple transitions between structures II (orange region, $M>0$) and H (blue region, $M<0$).
}
 \label{fig:Empty_LSMC_prod}
\end{figure}

Using this refined bias, we performed twenty new LSMC simulations, each of length \SI{5.44e8}{} steps. These runs produced good sampling over the whole o.p. range. Eighteen simulations exhibited several transitions between the two phases (see Fig.~\ref{fig:Empty_LSMC_prod}) while 2 runs showed only a single transition. Using eq.~\eqref{eq:dG} and retaining only the data of the 18 runs, we obtain 

\begin{equation}    \label{eq:deltaGammaEmpty42}
    \Delta \Gamma_\text{empty}^\text{II$\to$H} = 42.13 \pm 0.17\, k_{B}T. 
\end{equation}

This value is quite close to the initial estimate from the bias-generating TMMC simulation, indicating that the bias was already reasonably well converged. The production simulation allows a straightforward calculation of the uncertainty in $\Delta G$ by block analysis, with each production simulation treated as an independent block.

Notably, the value of $\Delta G$ obtained from the visited states histogram of the production simulations is insensitive to the binning used for the bias function. In contrast, estimates obtained directly from the bias function may depend weakly on the chosen binning. The good agreement between these two estimates therefore provides a useful consistency check that validates the reliability of our numerical result for $\Delta G_\text{empty}$.

\subsubsection*{Correction from the tails of the probability distributions}

The LSMC simulations described above sampled states over a restricted range of the order parameter and therefore did not fully cover the probability distribution $P(M)$ in either phase. Consequently, the left tail of $P(M)$ in sII and the right tail in sH were not included in the sampling.

To account for these missing contributions, we computed $P(M)$ separately within each phase using two isothermal–isobaric simulations. The resulting single-phase distributions were then stitched together with the LSMC data to reconstruct the full probability distribution $P(M)$. Re-evaluating the probability ratio in Eq.~\eqref{eq:dG} using this complete distribution yields the refined estimate $\Delta \Gamma_\text{empty}^{\text{II}\to\text{H}} \simeq 42.74\,k_B T$.

The correction arising from the neglected tails is therefore small. In practice, restricting the sampled order-parameter range to span the maxima of the probability peaks in the two phases is sufficient unless very high accuracy in $\Delta G$ is required.

\subsection{Free energy $\Delta G^{\text{II}\to\text{H}}$ for filled methane clathrates}
\label{sec:ResCH4}

We similarly computed, at $30$ bar and $294$\,K, the Gibbs free energy difference between methane clathrate structures II and H in the fully occupied case, where all cages contain a single CH$_4$ molecule. This quantity, $\Delta G_\text{s.o.}$, enters the thermodynamic cycle shown in Fig.~\ref{fig:cycle_filled}.

\begin{figure}[H]
 \centering
 \includegraphics[scale=0.45]{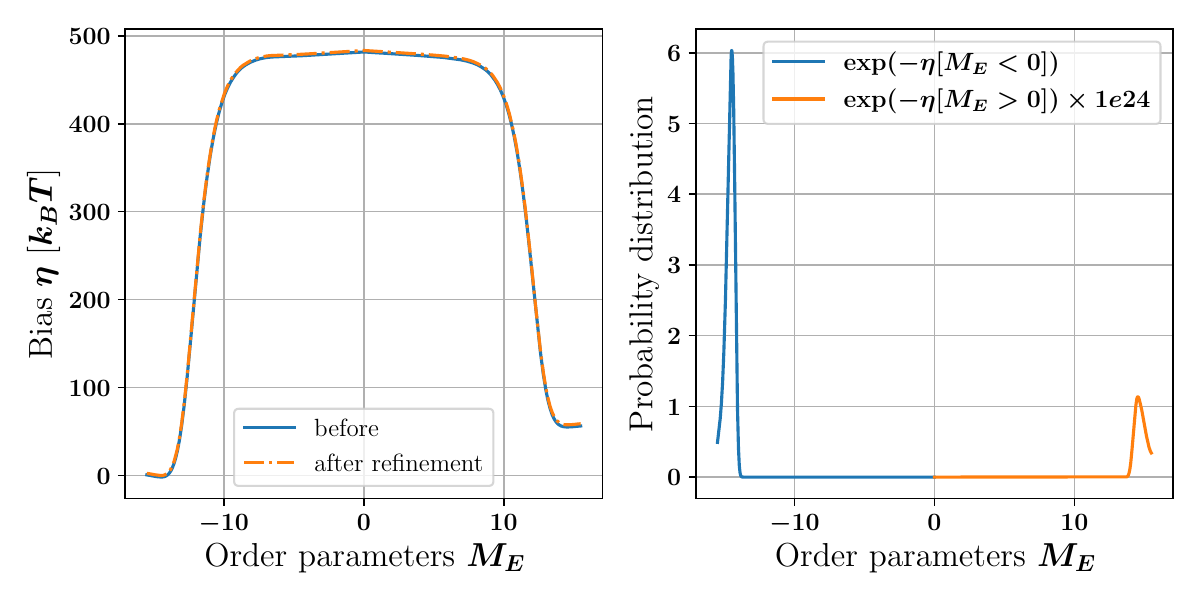}
 \caption{Bias function $\eta(M_E)$ (left) and corresponding probability distribution $\exp(-\eta(M_E))$ (right) for the singly-occupied methane hydrate at $T=294 K$ and $p=30$ bar. }
 \label{fgr:CH4_LSMC}
\end{figure}

Instead of using the o.p. defined in Eq.~\eqref{eq:M(sigma)}, which can exhibit large transient spikes (see Supplementary Information), we employed the logarithmic definition\cite{Graham2006}
\begin{equation}\label{eq:orderNew}
    M_{E} = \text{sgn}(E_\sigma - {E_\sigma'})\, \ln(1+|E_\sigma - {E_\sigma'}|),
\end{equation}
which compresses the range of the order parameter and improves sampling near gateway states. Single-phase $NPT$ simulations indicate that the bias function must be computed over the range $M_E \in [-15.5, 15.5]$. Since Transition-Matrix Monte Carlo is most efficient when the free energy profile is steep, we employed the Wang-Landau method to compute the bias inside the interval $[-9, 9]$ where the bias is relatively flat, and the transition-matrix method outside this interval. The resulting bias function is shown in Fig.~\ref{fgr:CH4_LSMC}. The free energy barrier is now higher at $\approx 480$ kJ/mol. Its ``top-hat'' shape is quite different than the sharp peak of $\eta(M)$ in Fig~\ref{fgr:Empty_LSMC}. This difference arises from the use of the logarithm. The bias function needs to be computed very accurately\footnote{In regions $[-15.5,-9]$ and $[9, 15.5]$, the transition-matrix simulations were run with 26 replicas working each on a narrow sub-interval of width 0.005 of the range. Length of each of these jobs: \SI{3.2e8} MC steps.}, with a maximal difference of about $1.5 k_BT$ between two adjacent bins, for a production LSMC simulation to explore the whole range of order parameters. The minima of the bias function, which correspond to probability peaks, are located around $M_E = \pm14.5$, i.e. near $M=\pm \SI{2e6}{}$ kJ/mol, similarly as for the empty clathrates. 

We used this bias function as input for 20 production LSMC simulations, each of length \SI{6.4e8}{} MC steps. The bias was subsequently refined using Eq.~\eqref{etaCorr}, and a further set of 20 production simulations was performed, each lasting \SI{4.2e8}{} MC steps.

\begin{figure}[htb]
 \centering
\includegraphics[width=0.75\linewidth]{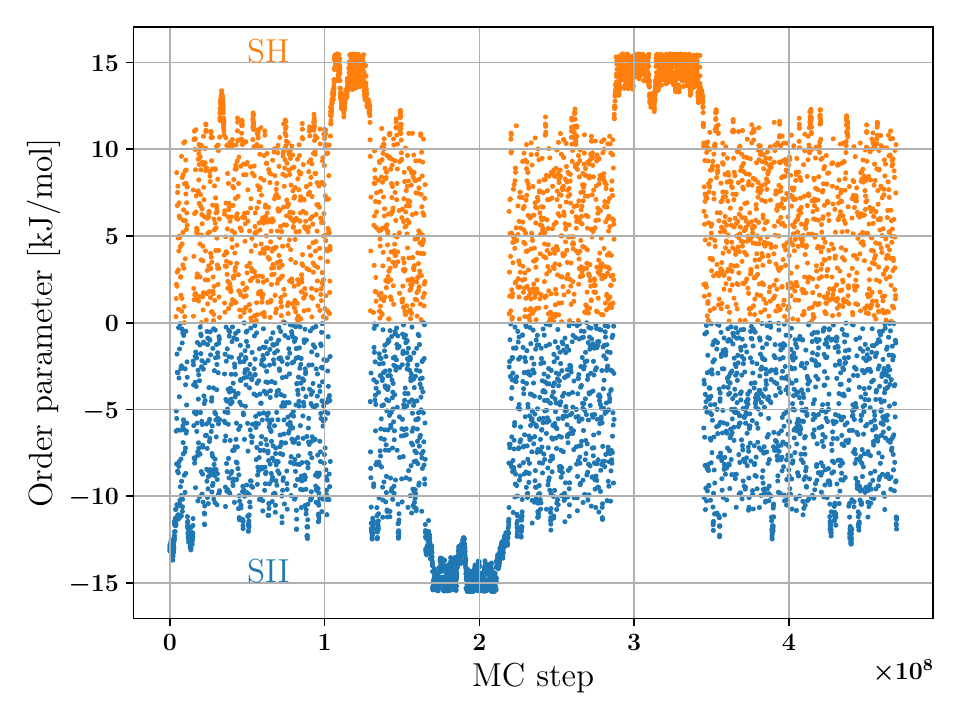}
\caption{
Logarithmic LS order parameter, Eq.~\eqref{eq:orderNew}, for a representative LSMC simulation of the fully occupied methane hydrate using the refined bias function shown in Fig.~\ref{fgr:CH4_LSMC}. Multiple transitions between phases II and H are observed.
}
 \label{fig:Filled_LSMC_prod}
\end{figure}

An example trajectory is shown in Fig.~\ref{fig:Filled_LSMC_prod}, where several transitions between phases II and H are observed. The simulations exhibit some difficulty in crossing the regions $[-12,\,-11]$ and $[11,\,12]$, indicating that the sampling is not perfectly flat; however, it remains sufficient for our purposes. 

Discarding three runs that did not adequately sample the characteristic states of both sII and sH, we obtain $\Delta G_\text{s.o.} = 57.23 \pm 0.36\, k_B T$.

\subsection{Correspondence between lattice sites}
\label{sct:correspondence}

The height of the free-energy barrier between the two phases depends on the degree of similarity between the corresponding lattices. We attempted to improve the mapping between lattice sites by matching sites associated with small cages in sII to those in sH and, where possible, preserving pairs of hydrogen-bonded water molecules. However, this strategy led to only marginal improvement, and we therefore retained the initial arbitrary mapping between lattice sites.

For filled hydrates, the lattice-switch move also requires a mapping between the reference lattice sites of the guest molecules. In principle, if a guest molecule were to hop between adjacent cages during the simulation, its reference lattice site would need to be updated on the fly. In practice, no such cage hopping was observed on the timescale of our simulations for fully occupied methane hydrates (see Fig.~\ref{fig:MSD} in the Supplementary Information), and thus no redefinition was required.

\subsection{Coexistence between structures II and H}
In this section, we calculate the various elements in the thermodynamic cycles used to compute the free energy difference $\Delta \Gamma$ between clathrate structures and to determine the coexistence pressure. Starting from Gibbs free energy differences obtained by Lattice-Switch Monte Carlo, we construct routes based on empty and fully occupied clathrates, and combine them with thermodynamic integration in $\mu_g$ and $P$ to obtain $\Delta \Gamma(P)$. The approach is illustrated for both argon and methane hydrates.

\subsubsection{Argon hydrate }
\label{sec:ResAr}
\label{sec:ThermoIntegResAr}

\begin{figure*}[h]
 \centering
    \includegraphics[width=7.5cm]{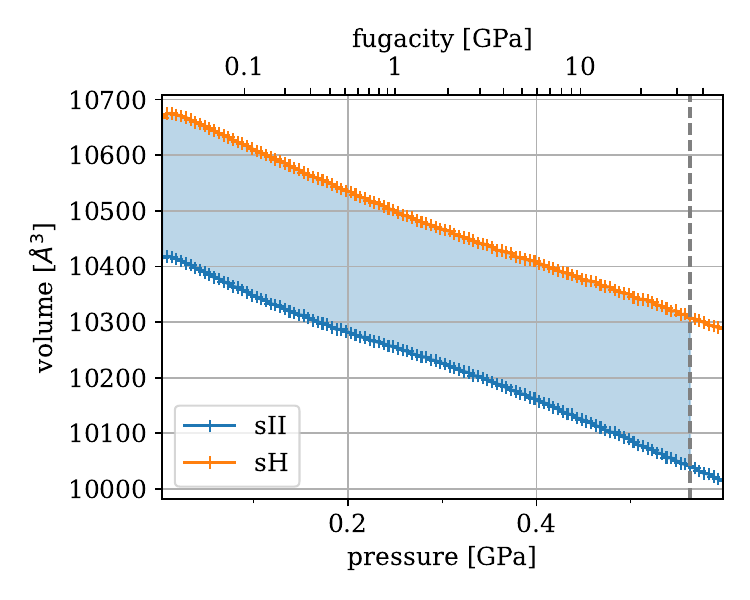}
    \qquad
   \includegraphics[width=7.5cm]{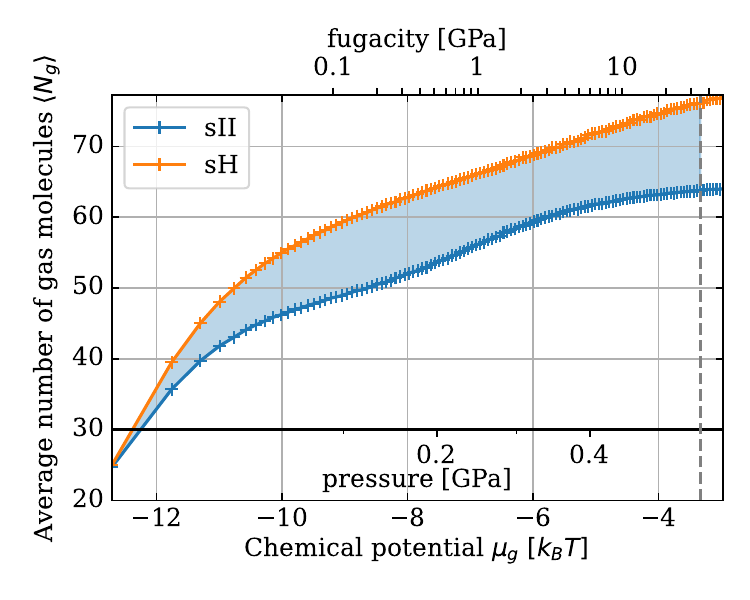}
\caption{
(Left) Volume of sII and sH argon hydrates as a function of argon pressure at 294~K; the corresponding argon fugacity is shown on the upper horizontal axis. (Right) Absorption isotherms for sII and sH argon hydrates as a function of the argon chemical potential, with the corresponding pressure shown on a secondary horizontal axis. The shaded regions represent the integrals $\int \langle V\rangle\, dP$ and $\int \langle N_g\rangle\, d\mu_g$ evaluated from 30 bar to the estimated coexistence pressure (term $\Delta \Gamma_{P \to P'}$).}

 \label{fgr:Ar_toP2}
\end{figure*}

\begin{figure}[h]
\centering
  \includegraphics[width=0.9\linewidth]{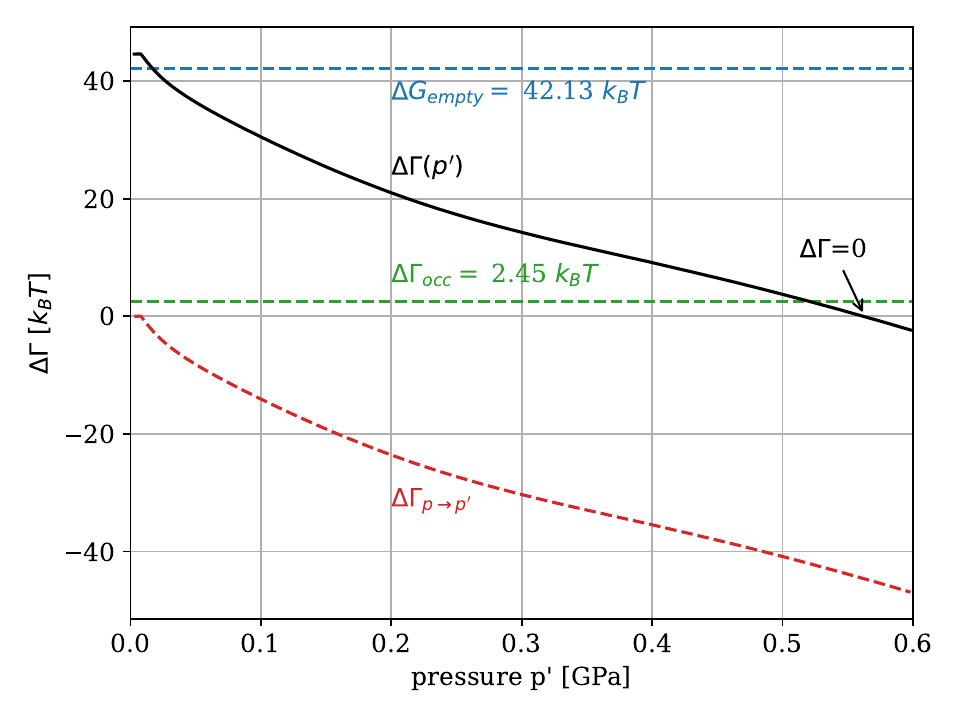}
  \caption{
  Free energy difference between argon sII and sH hydrates at 294~K calculated using $\Delta \Gamma(P) = \Delta \Gamma_{\text{empty}} + \Delta\Gamma_\text{occ} + \Delta \Gamma_{P \to P'}$. Coexistence ($\Delta \Gamma = 0$) occurs at 0.56 GPa.
  }
  \label{fgr:Ar_dGamma}
\end{figure}

For this example, we calculate the coexistence pressure at $294$K, following the thermodynamic cycle shown in Fig.~\ref{fig:cycle_simple}, which starts with an LSMC calculation for a system with empty cages. For this arrangement, the free energy difference $\Delta \Gamma_{\text{empty}}$ is known for a pressure $P=30$bar  (Eq.~\eqref{eq:deltaGammaEmpty42}).  

Following this cycle, the gas chemical potential—initially $\mu_g = -\infty$ for the empty hydrate—is first raised to the value corresponding to an applied gas pressure of 30 bar, namely $\mu_g = -12.71\, k_B T$, as given by the Lennard–Jones equation of state. 
%
%
To compute the associated free energy change $\Delta \Gamma_\text{occ}$ (Eq.~\eqref{eq:DeltaGammaOcc}) we performed constant-$(N_w \mu_g P T)$ simulations over the range $\mu_g \in [-25.5,\,-12.5]\,k_B T$. A total of 130 simulations were carried out for each structure at $P=30$ bar, each of length \SI{3.2e7}{} MC steps. Thermodynamic integration of $\langle N_g \rangle\, d\mu_g$ in each phase then yields (see SI, Fig.~\ref{fgr:Ar_Occupancy})
\begin{equation}
    \Delta \Gamma_\text{occ} = \Delta \Gamma_\text{occ}^\text{sH} - \Delta \Gamma_\text{occ}^\text{sII} = 2.44 \pm 0.04\, k_B T.
\end{equation}

Next, we compute the pressure dependence of $\Delta \Gamma$ using Eq.~\eqref{eq:DeltaGammaUse} and the thermodynamic cycle shown in Fig.~\ref{fig:cycle_coex}, with $\mu_g^\text{II} = \mu_g^\text{H}$. To this end, we performed 120 simulations in the $\Gamma$ ensemble for each phase over the pressure range $P \in [0.003,\,0.598]$ GPa, each of length \SI{3.2e7}{} MC steps, in order to evaluate the integrals $\int \langle V \rangle\, dP$ and $\int \langle N_g \rangle\, d\mu_g$ (see Fig.~\ref{fgr:Ar_toP2}).

In these simulations, the gas chemical potential $\mu_g(P,T)$ was adjusted according to the Lennard–Jones equation of state to maintain consistency between the imposed mechanical pressure and the gas pressure. The resulting free energy difference,
\begin{equation}    \label{eq:FormulaGammaForEmpty}
    \Delta \Gamma(P) = \Delta \Gamma_{\text{empty}} + \Delta \Gamma_\text{occ} + \Delta \Gamma_{P \to P'},
\end{equation}
and its individual contributions are shown in Fig.~\ref{fgr:Ar_dGamma}.

From this analysis, we estimate the coexistence pressure to be
\begin{equation}
    P_{\text{coex}}(\text{argon}) = 0.563 \pm 0.03\, \text{GPa},
\end{equation}
which is in reasonable agreement with the experimental value of 0.46 GPa reported for this transition.~\cite{Loveday2008}

\subsubsection{Methane hydrate}
\label{sec:ThermoIntegResCH4}

A cycle based on fully occupied clathrates (Fig.~\ref{fig:cycle_filled}), offers in principle a shorter integration path than that starting from empty structures. We have compared both approaches for estimating the coexistence pressure between sII and sH for a system of methane hydrates. Following both routes to the coexistence pressure provides a useful consistency check on our method and calculations.

\subsubsection*{Cycle starting from filled clathrates}

The Gibbs free energy difference between the filled cages-singly-occupied methane structures was calculated by LSMC to be $\Delta G_\text{s.o.} = 57.23\,k_B T$ (Sect.~\ref{sec:ResCH4}). For switching from the Gibbs to the $\Gamma$ free energy with eq.~\eqref{eq:GsoToGamma}, the gas chemical potential $\mu_g^{\alpha, \text{filled}}$ at which the average number of guest molecules is equal to the number of cages $\langle N_g \rangle = N_\text{cages} = 48$ needs to be determined. 

\begin{figure}[h]
\centering
  \includegraphics[width=\linewidth]{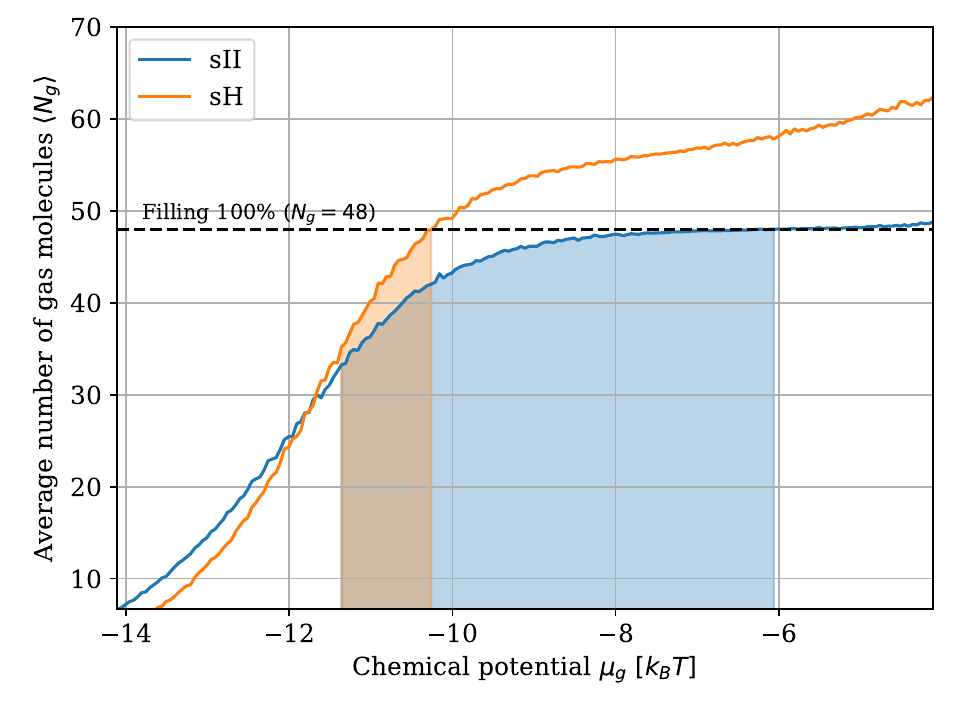}
  \caption{Average number of trapped methane molecules at $30$ bar and $294$ K as a function of the gas chemical potential from MC simulations in the $\Gamma$ ensemble. The shaded areas show the integral $\int \langle N_g \rangle \,d\mu_g$ between the state filled at 100\% (chemical potential $\mu_g^{\alpha, \text{*}}$ for $\alpha=\text{II or H}$) and the state at methane pressure 30~bar ($\mu_{g} = -11.37\,k_B T$), which provides $\Gamma_\text{occ}^\alpha$, the free energy change in the last step of the cycle in Fig~\ref{fig:cycle_filled}.
  }
  \label{fgr:CH4_Occupancy}
\end{figure}

Fig~\ref{fgr:CH4_Occupancy} shows absorption isotherms in the two structures computed at 30~bar using 200 Monte-Carlo simulations in the $\Gamma$ ensemble, from which one can estimate $\mu_g^{\text{sII}, *} \simeq -6.2\,k_B T$ and $\mu_g^{\text{sH}, *} \simeq -10.2 k_B T$ (* means `filled'). 
\footnote{As the absorption isotherm in sII is quite flat near full occupancy, an error on $\mu_g^{\text{sII}, *}$ can significantly impact an integral like $\int_{\mu_g^*}^{\mu_g'} \langle N_g \rangle \, d\mu_g$. Therefore, we refined the estimates for $\mu_g^{\alpha, *}$ by reweighting statistics on $P(N_g)$ obtained from a simulation with our initial values of $\mu_0 = \mu^{\alpha, *}_g$. By solving the reweighting equation
\begin{equation}
    \langle N \rangle_\mu
    = \frac{\langle N e^{(\mu - \mu_0) N }\rangle_{\mu_0}}{\langle e^{(\mu - \mu_0) N} \rangle_{\mu_0}}
    \qquad \text{where $N = N_g$}
\end{equation}
we found the chemical potential at which $\langle N_g \rangle = N_\text{cages}=48$, leading to the improved estimates
$$
\mu_g^{\text{II}, *} = -6.091\,k_B T
\qquad 
\text{and}
\qquad
\mu_g^{\text{H}, *} = -10.262 \,k_B T.
$$
As will be explained later, after eq.~\eqref{eq:Gamma30bar}, having very accurate estimates for these two chemical potentials is actually not strictly necessary because of a compensation that occurs in the free energy difference calculation
} 
Hence, the second term in eq.~\eqref{eq:EnsembleSwitch_Legendre} evaluates to $-(\mu_g^{H,*} - \mu_g^{II,*}) = 200.208\, k_B T$.
 

The second term in Eq.~\eqref{eq:GsoToGamma} involves $P(\eta = 1)$, the probability that all cages are singly occupied when $\langle N_g \rangle = N_\text{cages}$ (i.e.\ at the gas chemical potential $\mu_g^{\alpha,*}$). For sH, this probability is very small, since multiple occupancy of the large cages is readily possible, making the fully singly occupied configuration just one of many ways to achieve 100\% filling.

\begin{figure}[h]
    \centering
    \includegraphics[width=0.85\linewidth]{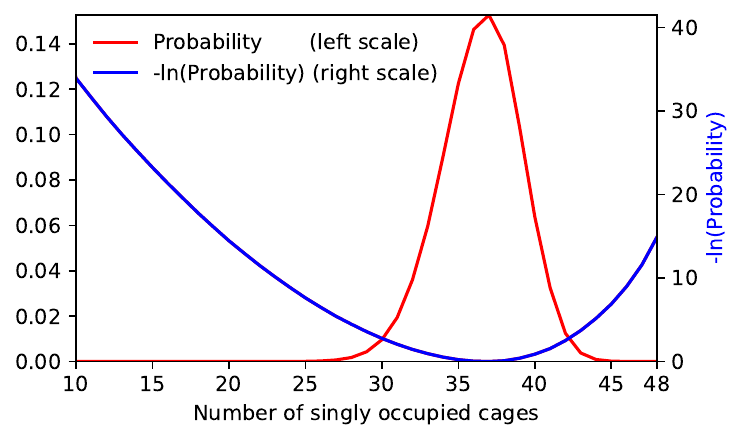}
    \caption{Probability distribution for the number of singly occupied cages in the case of a methane hydrate ($T = \SI{294}{K}$ and $P=\SI{30}{bar}$) with structure H made up of 4 unit cells (i.e. 48 cages).}
    \label{fig:Prob_single}
\end{figure}

To evaluate $P(\eta = 1)$, we performed simulations in the $\Gamma$ ensemble (lasting \SI{1.6e8}{} MC steps for sH and somewhat shorter for sII), employing flat sampling in the order parameter $\eta = N_g / N_\text{cages}$. The required bias function was computed using transition-matrix Monte Carlo, implemented within a custom version of \textsc{DL\_MONTE} adapted to this order parameter.
Using eq.~\eqref{eq:FluctCorr} together with the obtained distribution $P(\eta)$ (Fig.~\ref{fig:Prob_single}), we obtain $\Delta \Gamma^\text{II}_\text{s.o.} = 0.081\,k_B T$ and $\Delta \Gamma^\text{H}_\text{s.o.} = 16.825\,k_B T$. The corresponding probability for full single occupancy in sH $P(\eta = 1) = \exp(-\beta \Delta\Gamma^\text{H}_\text{s.o.})= \SI{4.9e-8}{}$ is extremely small. The large value of $\Delta \Gamma^\text{H}_\text{s.o.}$ reflects the substantial entropy difference between the constrained state, in which all cages are singly occupied, and the unconstrained state.

Moving on to the second row of the cycle in Fig.~\ref{fig:cycle_filled}, this corresponds to a free energy change
\begin{equation}
    \Delta(G_\text{s.o.} \to \Gamma)  = 200.208 - (16.825 - 0.081) = 183.464\, k_B T.
\end{equation}

Finally, concerning the third row in Fig.~\ref{fig:cycle_filled}, we 
compute $\Delta\Gamma_\text{occ}^\alpha$ ($\alpha = \text{II or H})$ by integrating $- \langle N_g\rangle  d\mu_g$ from the chemical potential $\mu_g^{*}$ (full occupancy) to the chemical potential of the gas at our reference pressure and temperature, namely $\mu_g(\text{30 bar},\, \text{294\,K}) = -11.37\, k_B T$ (LJ equation of state). $\Delta\Gamma_\text{occ}^\alpha$ corresponds to the shaded areas under the isotherms in Fig.~\ref{fgr:CH4_Occupancy}. This yields
\begin{equation}    \label{eq:GammaOccMethane}
    \Delta\Gamma_\text{occ} = \Delta\Gamma_\text{occ}^\text{H} - \Delta\Gamma_\text{occ}^\text{II}
    = 46.72 - 236.56 = - 189.84 \pm0.04\, k_B T. 
\end{equation}
Collecting these results together, the free energy difference between sII and sH with cage occupancies corresponding to the initial methane pressure is found to be
\begin{equation}    \label{eq:Gamma30bar}
    \Delta\Gamma(\text{30 bar}) = \Delta G_\text{s.o.} + \Delta(G_\text{s.o.} \to \Gamma) + \Delta\Gamma_\text{occ} = 50.85 \pm 0.36 \, k_B T
\end{equation}

The strong cancellation between Eqs.~\eqref{eq:GammaOccMethane} and \eqref{eq:GammaOccMethane} originates from the similarity between the shaded areas under the isotherms in Fig.~\ref{fgr:CH4_Occupancy} and the area under the line $N_g=48$ over the same interval of $\mu_g$. As a result, any imprecision in the integration bound $\mu_g^{\alpha,*}$ has only a minor effect on the final result~\eqref{eq:Gamma30bar}, particularly when $\langle N_g \rangle$ is nearly flat in the vicinity of $\mu_g^{\alpha,*}$.

In fact, the rectangular contribution $(\mu_g^{\alpha,*} - \mu_g(\text{30 bar})) \times \min(\langle N_g \rangle)$ cancels exactly between the terms $\Delta(G^\alpha_\text{s.o.} \to \Gamma^\alpha)$ and $\Delta \Gamma^\alpha_\text{occ}$. This observation can be exploited to redefine these terms by subtracting this common contribution, thereby eliminating two large compensating terms from the calculation.

\begin{figure}[htb]
 \centering
 \begin{subfigure}[t]{\textwidth}
         \caption{From empty clathrates}
     \includegraphics[width=0.45\linewidth]{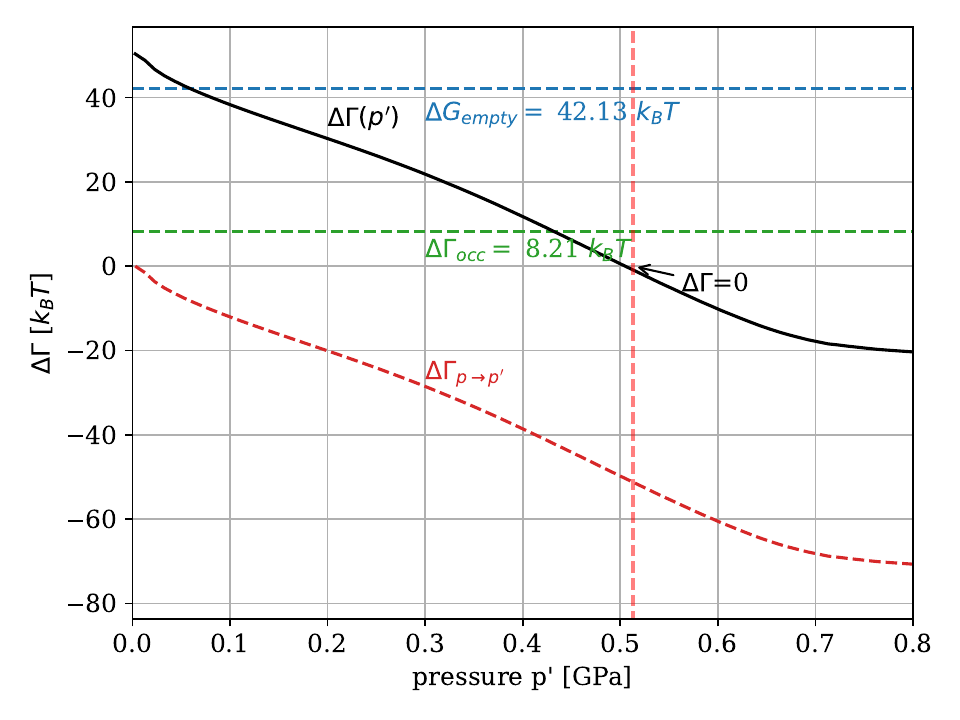}
 \end{subfigure}
 \begin{subfigure}[t]{\textwidth}
        \caption{From filled clathrates}
    \includegraphics[width=0.45\linewidth]{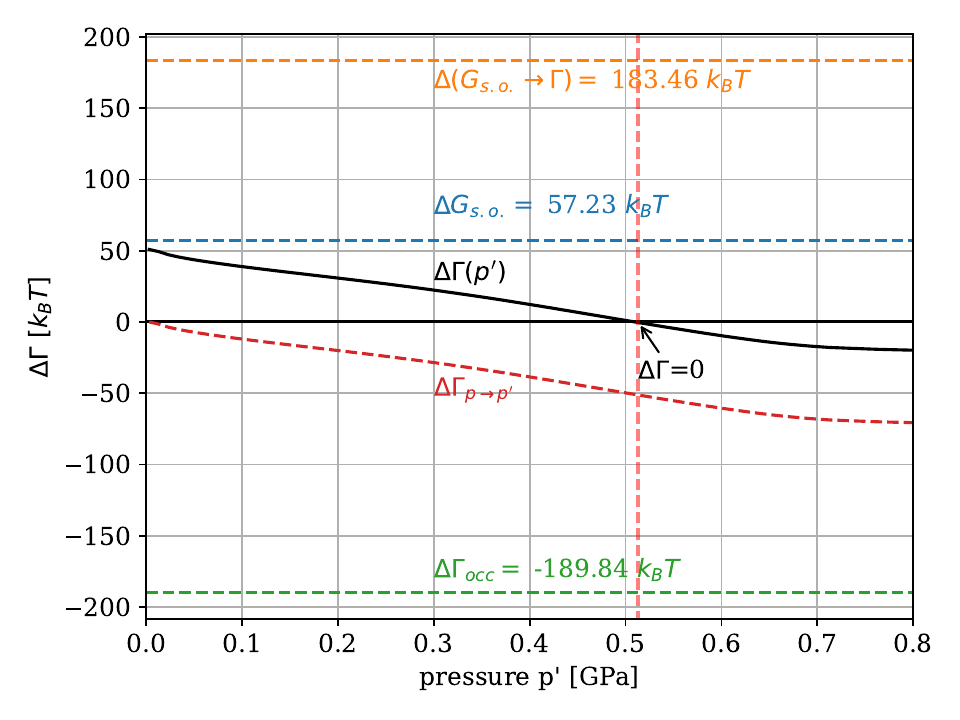}
 \end{subfigure}
 \caption{
The free energy difference $\Delta \Gamma(P)$ between methane sII and sH hydrates at 294~K, computed via two independent routes: (a) starting from empty clathrates using Eq.~\eqref{eq:FormulaGammaForEmpty}, and (b) starting from filled clathrates using
$
\Delta \Gamma(P) = \Delta G_\text{s.o.} + \Delta(G_\text{s.o.} \to \Gamma) + \Delta \Gamma_\text{occ} + \Delta \Gamma_{P \to P'}.
$
Both approaches yield consistent results, with phase coexistence occurring at $P \simeq 0.51$~GPa. The strong cancellation between two contributions in panel (b) can be avoided by a suitable redefinition of these terms (see main text).
}
  \label{fgr:CH4_dGamma}
\end{figure}

Next, we compute the pressure dependence of the free energy, $\Delta \Gamma_{P \to P'}$, by performing 100 simulations in the constant-$(N_w \mu_g P' T)$ ensemble, with $\mu_g = \mu_g(P',T)$, over the range $P' \in [0.003,\,0.993]$ GPa, using Eq.~\eqref{eq:DeltaGammaUse}. The corresponding variations of the volume and cage occupancies with pressure are shown in Fig.~\ref{fgr:CH4_toP2} of the SI, while the resulting $\Delta \Gamma_{P \to P'}$ is plotted in Fig.~\ref{fgr:CH4_dGamma}.

From this figure, the coexistence pressure is identified as the value of $P'$ for which $\Delta \Gamma(P')$ vanishes. We obtain, in our model,
\begin{equation}
\label{eq:PcoexMethane}
    P_\text{coex}(\text{methane}) = 0.513 \pm 0.005\,\text{GPa}.
\end{equation}

%

\subsubsection*{Cycle starting from empty clathrates}

We have also estimated the coexistence pressure starting from empty clathrates, following the same procedure as for argon hydrate. The resulting free energy $\Delta \Gamma(P)$, together with its individual contributions, is shown in Fig.~\ref{fgr:CH4_dGamma}. This curve agrees very well with that obtained when starting from filled clathrates.

In particular, at 30 bar we find $\Delta \Gamma = 50.34 \pm 0.18\, k_B T$, in excellent agreement with the previous result~\eqref{eq:Gamma30bar}. The same coexistence pressure, Eq.~\eqref{eq:PcoexMethane}, is also recovered.
\bigskip

Coexistence between sII and sH methane hydrates has been observed experimentally\cite{Loveday2008,Chou2000, Choukroun2007, Shu2011}. Methane sII is expected to be metastable with respect to sI.~\cite{Loveday2008}. Since the transition between methane sI and sH occurs at $\sim 0.8$\,GPa\cite{Hirai2015, Loveday2008}c, the less stable sII is expected to transform into sH at a lower pressure. Shu et al.\cite{Shu2011} reported that sH is stable from 0.6 to 0.9\,GPa and that it can coexist with sII at 0.6\,GPa. The predicted pressure~\eqref{eq:PcoexMethane} appears as plausible. 

When comparing with experiments, it should be noted that the simulations correspond to slightly different conditions (presence of excess gas rather than excess water), and that the semi-empirical interaction potentials employed here are not specifically optimised for high-pressure regimes.

\section{Conclusions}
\label{sec:Ccl}
 
We have shown that Lattice-Switch Monte Carlo, originally developed as an efficient method for computing free energy differences in simple atomic crystals, can be extended to significantly more complex systems such as gas hydrates. Its main advantages are its directness (it follows a path that connects the two competing phases directly), its ability to provide well-defined statistical uncertainties, and its applicability to both filled and empty clathrate structures at a prescribed mechanical pressure.
The free energy profile (barrier) along the lattice-switch path can be determined using efficient techniques such as transition-matrix Monte Carlo or the Wang–Landau algorithm. Although this barrier is substantially higher in gas hydrates than in simple atomic systems, it can still be overcome within LSMC simulations. The overall computational cost of LSMC scales linearly with the number of molecules.


Because distinct clathrate structures generally have different compositions, their relative thermodynamic stability cannot be assessed solely from their Gibbs free energies. For gas hydrates in equilibrium with a gas reservoir, it is more convenient to formulate the stability criterion in terms of the thermodynamic potential
$\Gamma(N_w,\mu_g,P,T) = G - N_g \mu_g = N_w \mu_w.$ Working in the constant-$(N_w,\mu_g,P,T)$ ensemble is therefore particularly well suited to this problem. We have detailed the statistical-mechanical properties of this relatively little-used ensemble, which is employed extensively throughout this work.

Using thermodynamic cycles, we have computed the free energy difference $\Delta \Gamma$ between clathrate structures II and H, starting either from empty (Fig.~\ref{fig:cycle_simple}) or fully occupied clathrates (Fig.~\ref{fig:cycle_filled}). Both approaches yield consistent results, leading to the same $\Delta \Gamma(P)$ and coexistence pressure. 

This consistency is obtained only when the free energy difference between the fully occupied state (all cages singly occupied) and the unconstrained occupancy state is properly accounted for via Eq.~\eqref{eq:FluctCorr}. This contribution, which is predominantly entropic, can be substantial when multiple occupancy of cages is possible, as in the large cages of sH.  For example, in methane sH hydrate at $30\,$bar, we find this contribution to be significant, of the order of $\sim 17\,k_B T$. 

The cycle starting from empty clathrates is simpler to implement, however empty clathrates are metastable and significantly different from the filled hydrates of interest. In our simulations, the empty structure~H was stable at 30 bar and 294\,K over the simulation time scale, but it may collapse at higher pressures (the empty structures I and II , which contain smaller voids, are expected to be more stable under pressure, especially if the temperature is low~\cite{Falenty2014, Cruz2019}). In contrast, the cycle starting from filled clathrates follows a shorter thermodynamic path\footnote{The path could be further shortened by performing the lattice-switch simulation at the pressure at which $\mu_g(P,T) = \mu_g^{\alpha, *}$.} to the states of interest, which can improve accuracy. This approach requires fewer absorption simulations but is more specialized, as it depends on the specific guest molecule and it requires two additional calculations for each structure, namely the determination of the gas chemical potential $\mu_g^{\alpha, *}$ that yields a cage filling of 100\% and the probability $P(\eta = 1)$. Overall, when high accuracy is not essential and metastability is not a concern, the cycle starting from empty clathrates remains the preferred approach.

To our knowledge, the exact calculation of the coexistence pressure between two clathrate structures for a given interaction model—without neglecting entropic contributions (e.g. those arising from phonons and cage-occupancy disorder)—has not been reported previously. The coexistence pressures obtained here for sII–sH transitions in argon and methane hydrates are in overall good agreement with available experimental data.

Possible extensions of this work include determining coexistence at fixed overall composition, where phase separation into, for example, hydrate and pure ice may occur, and computing free energy differences involving other structures such as sI and pure ice phases. These directions are left for future study.

\section*{Author contributions}
All authors were involved in conceiving and directing the research. OM performed the numerical calculations. 

\section*{Conflicts of interest}
There are no conflicts to declare.

\section*{Data availability}

The data underlying this study are available in the published article and its Supporting Information (SI). The SI contains additional plots, details and input files for the simulation program DL\_MONTE~\cite{dlmontesource} version 2.08, including the refined bias functions used in the production LSMC simulations of the empty and filled methane hydrates.

\section*{Acknowledgements}
The authors are grateful to Tom Underwood for useful discussions and for help with DL\_MONTE, in particular with implementing a custom order parameter in free energy calculations. The computer simulations were carried out using the computational facilities of the Institute UTINAM and the computational facilities of the Advanced Computing Research Centre, University of Bristol, funded by EPSRC (EP/T022078/1). This work has been supported by the EIPHI Graduate School (contract ANR-17-EURE-0002) and by the Bourgogne-Franche-Comté Region. OM was supported by the School of Physics, University of Bristol.




\balance

\bibliography{rsc} 
\bibliographystyle{rsc} 

\appendix

\section{Upper bound for the free energy of lifting the single-occupancy constraint}
\label{AppendixA}

One can establish an upper bound, \(\Delta\Gamma_{\text{s.o.}}^+\), for \(\Delta\Gamma_{\text{s.o.}}\), under the assumption that interactions between guest molecules increase the probability of single occupancy compared to the ideal case, where guest molecules are non-interacting. In other words, interactions energetically disfavor multiple guest molecules occupying the same cell. 

To derive this bound, we consider the scenario in which the guest particles are ideal. Using combinatorial reasoning, we analyze the unrestricted assignment of $N_g$ non-interacting guest molecules to $M$ cells under the condition $N_g = M$. In this idealized case, the probability of finding $k$ guest molecules in a given cell is given by:

\begin{equation}
P_{occ}(k)={M\choose k}\left(\frac{1}{M}\right)^k\left(1-\frac{1}{M}\right)^{M-k}
\label{eq:Pocc}
\end{equation}
which for large $M$ is Poissonian. Since all cells are equivalent, this is also the distribution for any cell.

From this one deduces the probability of all $M$ cells being singly occupied:

\begin{equation}
P(\eta=1)=[P_{\text{occ}}(k=1)]^M=\frac{M!}{M^M}.
\end{equation}

Thus 

$$
\Delta\Gamma_{\text{s.o.}}^+=-k_B T\, \ln\left(\frac{M!}{M^M}\right)\approx k_B T\, M +O(\ln M)
$$
which is extensive for large $M$, as required thermodynamically. 

For methane guest molecules, we find the equilibrium fraction of singly occupied calls to be $\eta^\star\approx 0.75$ (cf. Fig.~\ref{fig:Prob_single},  $\eta^\star$ is the equilibrium value of $\eta$ given $\langle N_g \rangle =48$) compared to $\eta^\star\approx 0.372$ for an ideal gas as calculated from Eq.~\ref{eq:Pocc}. This indicates that the probability of a cell being singly occupied is indeed considerably greater in the real system than in the ideal gas. Accordingly the measured value of $\Delta\Gamma_{\text{s.o.}}=16.83\,k_BT< \Delta\Gamma_{\text{s.o.}}^+\approx 48\,k_BT$.

\newpage

\onecolumn
\centerline{\textbf{\huge Supplementary information}}
\setcounter{figure}{0}
\setcounter{section}{0}
\makeatletter
\renewcommand \thefigure{S\@arabic\c@figure}
\makeatother

\bigskip

\begin{figure}[h]
\centerline{
\includegraphics[scale=0.35]{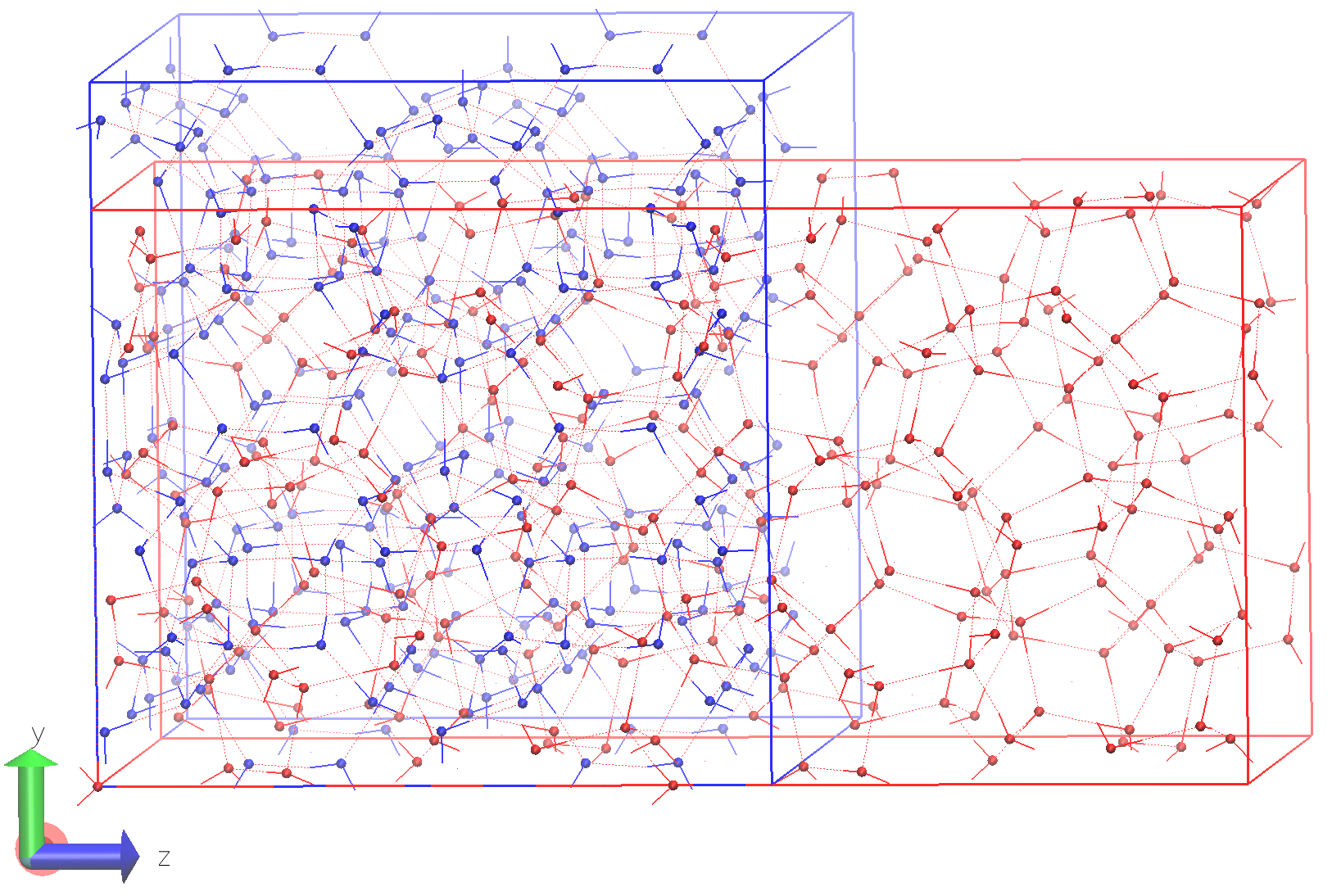}
}
\caption{\label{figBoxes}
Two unit cells of an empty clathrate structure II  (red orthorhombic box) alongside 8 unit cells with structure H (blue orthorhombic box). The blue box is almost cubic with cell dimensions $L_x = 2.442,\,  L_y=2.115$ and $L_z=2.028$~nm. Each box contains 272 water molecules. Thin dashed lines represent hydrogen bonds.
}
\end{figure}

The simulation input files of our production Lattice-Switch Monte Carlo simulations of empty and filled methane hydrates using the simulation program DL\_MONTE are {available at \url{XXX}{}}. Snapshots of the simulation boxes of empty clathrates with structures II and H are shown in Fig.~\ref{figBoxes}. The simulations were performed at the temperature 294\,.15K (number rounded to 294~K in the text of the article).

\section{Additional plots}

\begin{figure}[H]
    \centering
    \includegraphics[width=0.5\linewidth]{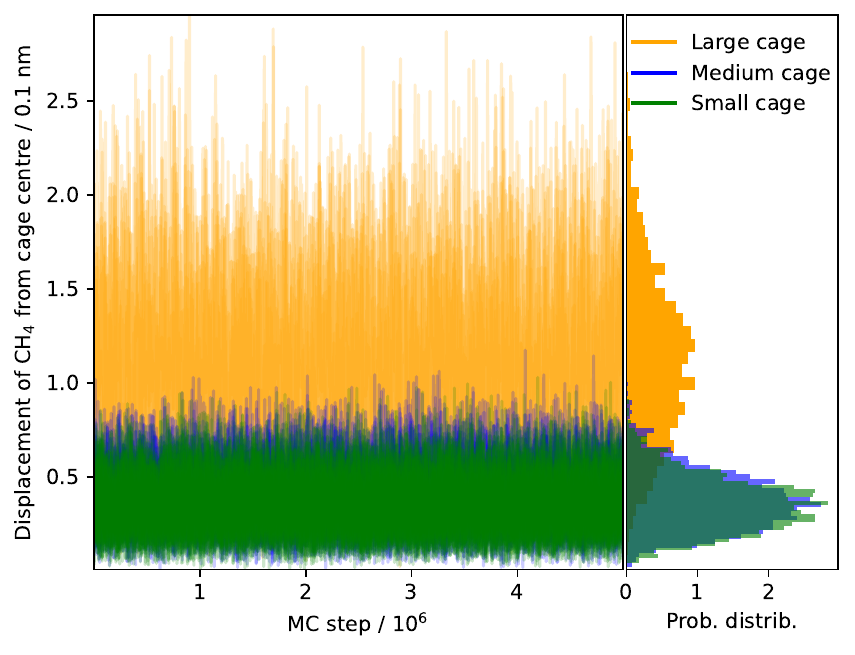}
\caption{
Norm of the displacement of each guest molecule from the center of its cage in a fully occupied methane sH clathrate at 30~bar and 294~K. The right panel shows the corresponding normalised histograms. Molecules are rarely found very close to the cage center, partly due to a geometric effect: the volume element $4\pi r^2\,dr$ decreases as $r \to 0$. No cage hopping is observed on the simulated timescale.
}
    \label{fig:MSD}
\end{figure}

\begin{figure}[H]
\centering
  \includegraphics[width=0.4\linewidth]{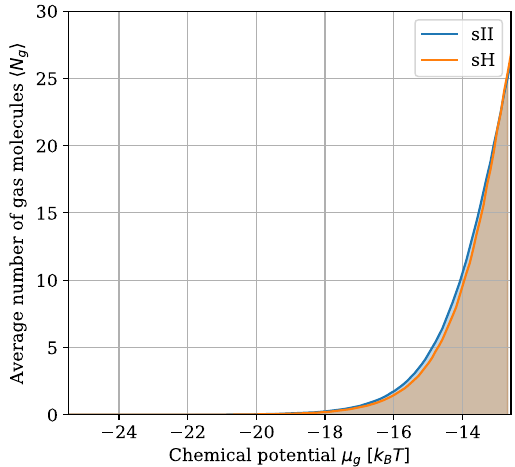}
  \caption{Average number of argon atoms in a hydrate with structure $\alpha = \text{II or H}$ as a function of the gas chemical potential (mechanical stress $P=30$ bar, $T=294$~K). The number of cages is 48.
  The shaded area shows the free energy $\Delta\Gamma_\text{occ}^\alpha$ of filling the cages.
 }
  \label{fgr:Ar_Occupancy}
\end{figure}

\begin{figure}[h]
 \centering
    \includegraphics[width=9cm]{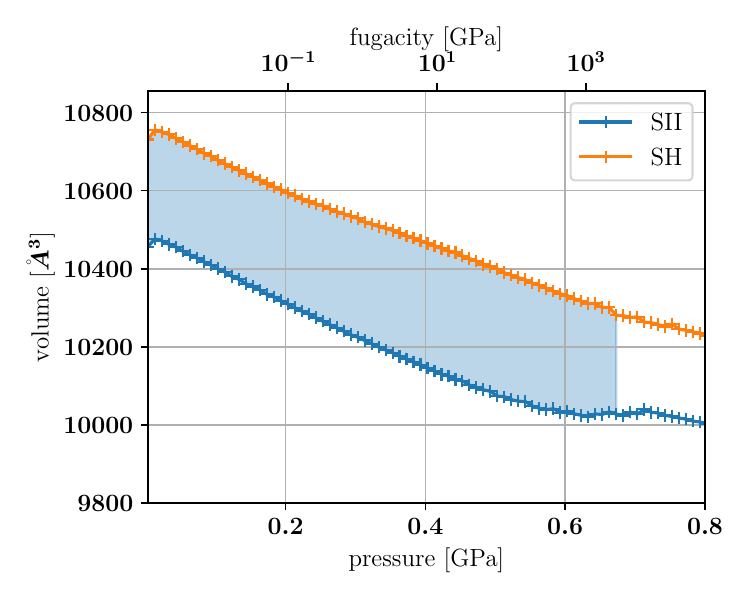}
   \includegraphics[width=9cm]{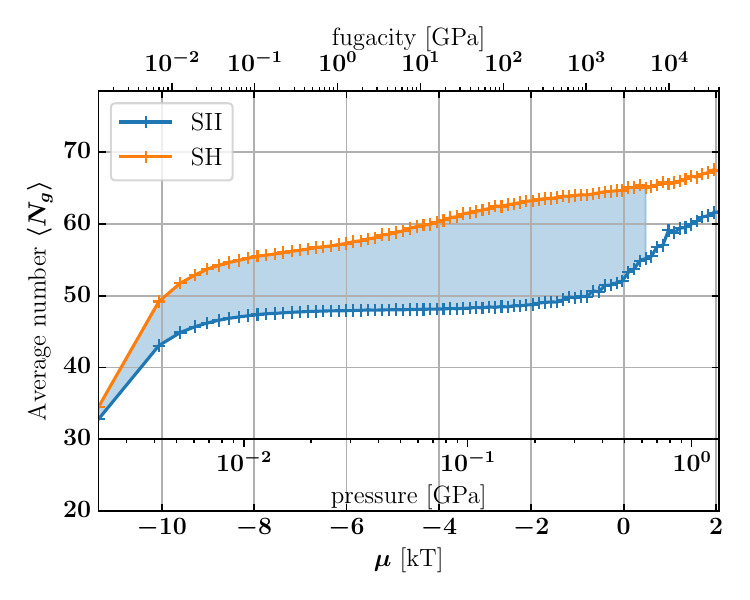}
 \caption{(Left) Volume of sII and sH methane hydrates as a function of methane pressure (at 294~K). The methane fugacity is indicated on the upper horizontal axis. (Right) absorption isotherms in sII and sH methane hydrates in terms of the methane chemical potential (corresponding pressure shown on a second horizontal axis). The shaded areas show the integral of $\langle V\rangle dP$ and $\langle N_g \rangle d\mu_g$ from 30 bar to the estimated coexistence pressure.}
 \label{fgr:CH4_toP2}
\end{figure}

\section{Spikes in the Lattice-Switch order parameter}	\label{sec spikes}

\begin{figure}[h]
\centerline{
\includegraphics[scale=0.55]{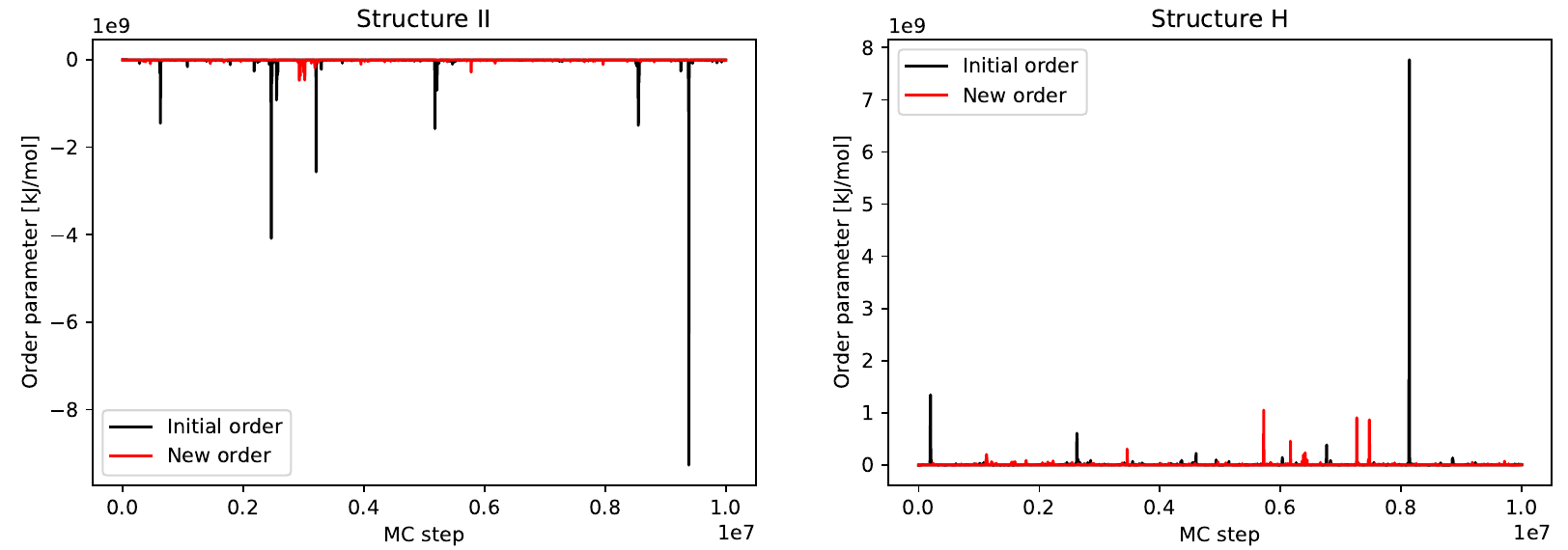}
}
\caption{\label{figOP}
Lattice-Switch order parameter [eq.~\eqref{eq:M(sigma)} in the main article] during an isothermal-isobaric simulation of an empty clathrate with structure II (left plot) or structure H (right plot) at 294~K and 30 bar. The curve labeled "New order" represents an alternative mapping choice between the lattice sites of structures II and H, as explained in the main text.
}
\end{figure}

Spikes (very short-lived large values) of the lattice-switch order parameter have been observed in our simulations of both the empty and filled clathrates. This phenomenon seems to occur more frequently in the empty clathrate. Fig.~\ref{figOP} shows such spikes during constant-$NPT$ simulations of an empty clathrate. The order parameter is computed for a tentative lattice-switch between structures II and H, but no such switch is actually performed.

The order parameter can attain extremely large values during spikes, reaching up to $\sim 10^{10}$ kJ/mol, whereas it typically fluctuates around $\sim \SI{2e6}{}$ kJ/mol under normal conditions.
In a lattice-switch simulation, each site in one structure is mapped onto a corresponding site in the other. In the present case, this mapping was initially chosen in an essentially arbitrary manner, as described in the main article. Analysis of configurations associated with these spikes shows that they originate from pairs of molecules becoming unrealistically close under a (virtual) lattice-switch move. Although the move itself is not accepted, it is used to evaluate the energy difference. Two water molecules that are well separated in structure II may be mapped onto neighbouring lattice sites in structure H. Since these molecules interact only weakly in the original structure, their displacements relative to their reference sites are largely uncorrelated. When mapped onto adjacent sites, they can therefore overlap or come into very close proximity, generating large energy differences and hence spikes in the order parameter. This effect is particularly pronounced near the boundaries of the simulation box, where periodic boundary conditions can bring mapped neighbours into close contact. It is also more frequent in empty hydrates, where the water network is less constrained and exhibits larger displacements from lattice sites than in filled clathrates.

\begin{figure}[h]
\centerline{\includegraphics[scale=0.44]{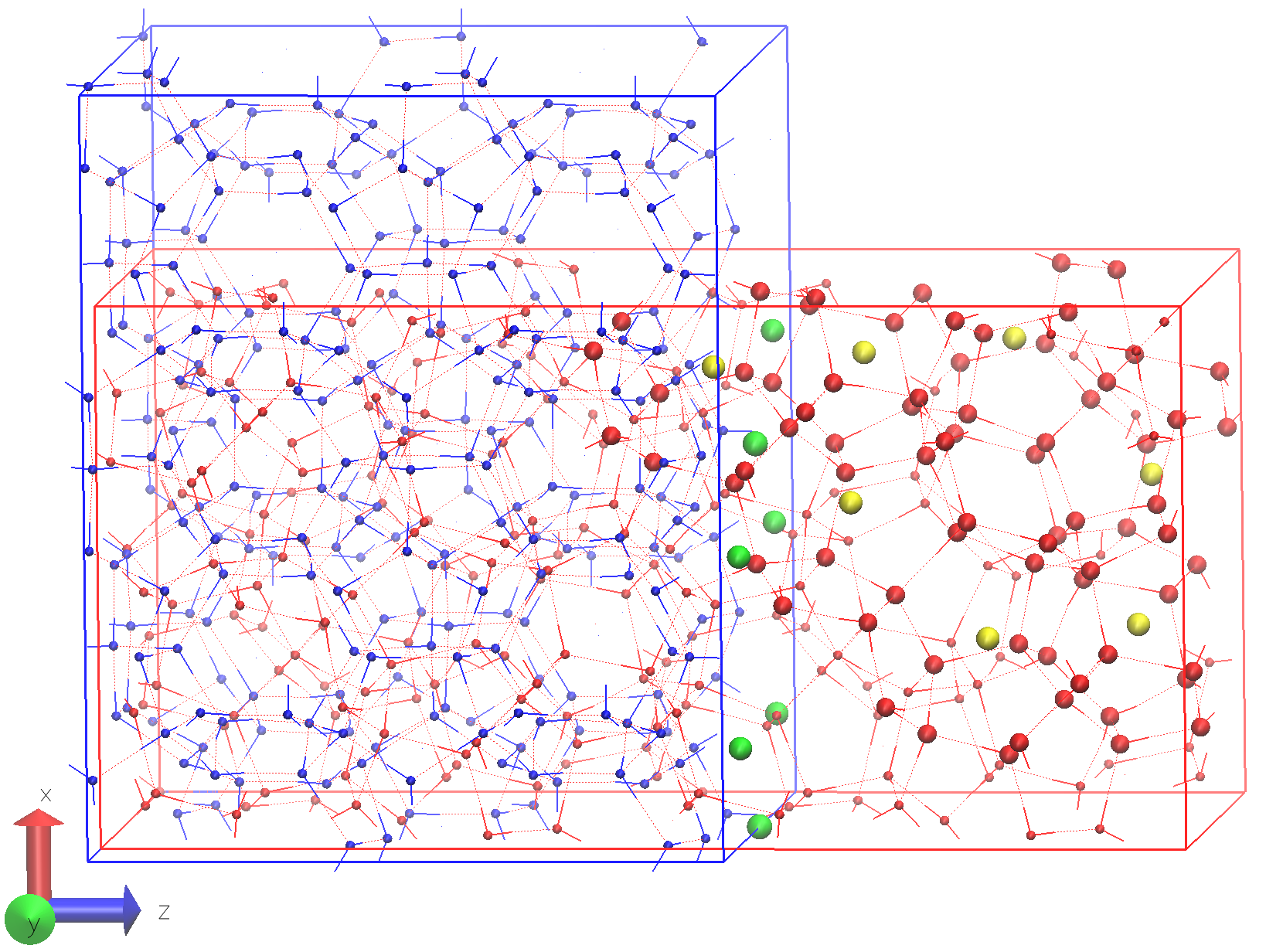}}
\centerline{\includegraphics[scale=0.44]{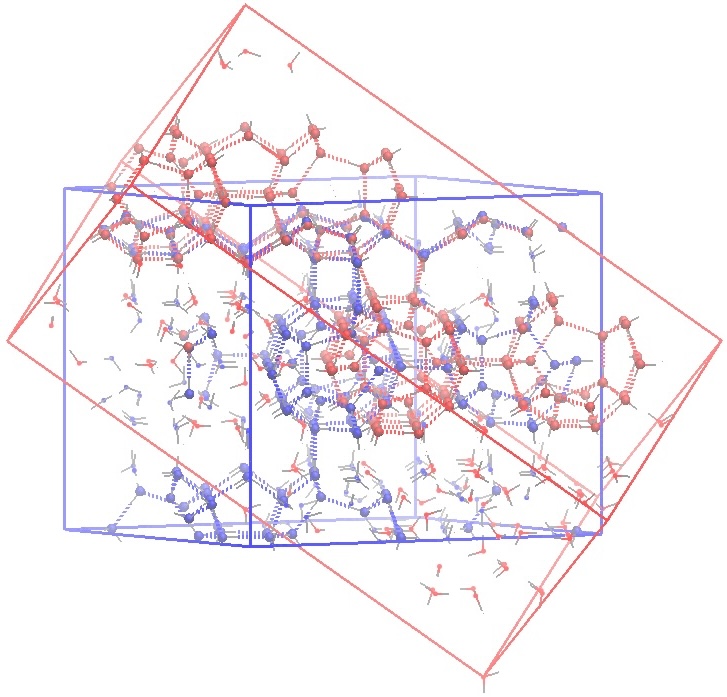}

}
\caption{\label{figLayer1}
(top) Same as Fig.~\ref{figBoxes} but where 7 small cages in sII forming a layer are marked with larger beads and by locating their cage centre with yellow spheres. 7 green spheres mark the corresponding cage centres in sH. \\
(bottom) A view showing some correspondences between lattice sites in sII (rotated and shifted red box) and sH (blue box).
}
\end{figure}

We attempted to construct a more physically motivated mapping between structures II and H by exploiting their shared structural features. In particular, both structures contain identical small $5^{12}$ cages. We therefore began by mapping the group of five small cages in the upper part of structure H (Fig.~2 of Ref.~\cite{HydrateReview}) onto the corresponding group in sII. In practice, the layer containing this group comprises seven small cages that could, in principle, be matched to sH (see Fig.~\ref{figLayer1}), allowing us to identify a consistent mapping for 98 out of the 272 water molecules in the simulation box.

Further correspondences between small cages can be established up to small shifts (see the central cages in Fig.~\ref{figLayer1}). However, identifying these mappings is non-trivial: the combination of rotations (Fig.~\ref{figLayer1}, bottom) and periodic boundary conditions in relatively small simulation boxes introduces ambiguities. In particular, when attempting to map layers of small cages in the $xy$-plane of sH onto those in sII, some molecules expected to belong to a given layer are reassigned to neighbouring layers once periodic boundary conditions are applied to the tilted sII simulation box.

Additional water molecules in the vicinity of the small cages were mapped by following the hydrogen-bond network as closely as possible. The remaining molecules were assigned by superimposing the two simulation boxes (Fig.~\ref{figLayer1}, bottom) and applying a nearest-distance criterion under periodic boundary conditions. This procedure yields a mapping between structures II and H that is somewhat more optimised, although not unique.

When using this new mapping, spikes are reduced in magnitude and occur less frequently in structure II, as illustrated by the curve labelled ``new order'' in Fig.~\ref{figOP}. In structure H, the frequency of spikes appears largely unchanged, although the very large spikes observed previously are no longer present. Despite this, the improvement in the overall accuracy of the LSMC simulations is marginal. A new bias function constructed with this mapping yields a free-energy barrier to the gateway states that is essentially unchanged, indicating that the intrinsic difficulty of the simulation is unaffected. For this reason, we retained the original, essentially random, mapping between the two structures. We also note that the use of the logarithmic order parameter, Eq.~\eqref{eq:orderNew}, mitigates the impact of large excursions in the order parameter. Since the spikes are transient and contribute negligibly to the free energy, their effect on the final results is minimal.
\end{document}